\documentclass[useAMS,usenatbib]{mn2e}
\usepackage{times}
\usepackage{graphicx, latexsym, amssymb, amscd, psfrag}
\usepackage{epsfig}
\usepackage{color}

\title[Star formation and quenching across the Coma supercluster]{Star
  formation, starbursts and quenching across the Coma supercluster}
\author[Mahajan, Haines and Raychaudhury]{Smriti
  Mahajan\thanks{e-mail:
    sm@star.sr.bham.ac.uk}, Chris P. Haines, Somak Raychaudhury\\
  School of Physics and Astronomy, University of Birmingham,
  Birmingham B15~2TT, UK}

\def\m{{$\mu$m}}
\def\eg{{e.g.}}
\def\ie{{i.e.}}
\newcommand{\kmsmpc}{~km~s$^{-1}$ Mpc$^{-1}$}

\begin{document}

\date{}

\pagerange{\pageref{firstpage}--\pageref{lastpage}} \pubyear{2010}

%\draft
%\special{!userdict begin /bop-hook{gsave 200 30 translate 65 rotate
%/Times-Roman findfont 216 scalefont setfont 0 0 moveto 0.93 setgray
%(DRAFT) show grestore}def end}

\maketitle

\label{firstpage}
%======================== ABSTRACT ==================================
 \begin{abstract}
 
   We analyse Spitzer MIPS 24$\mu$m observations, and Sloan Digital Sky Survey 
   (DR7) optical broadband photometry and spectra, to
   investigate the star formation properties of galaxies residing in
   the Coma supercluster region.  We find that star formation (SF) in dwarf
   galaxies is quenched only in the high density environment at the
   centre of clusters and groups, but that passively-evolving massive
   galaxies are found in all environments, indicating that massive
   galaxies can become passive via internal processes. The SF-density
   relation observed for the massive galaxies is weaker relative to
   the dwarfs, but both show a trend for the fraction of star-forming
   galaxies ($f_{SF}$) declining to $\sim\!0$ in the cluster cores. We
   find AGN activity is also suppressed among massive galaxies
   residing in the cluster cores.

   We present evidence for a strong dependence of the mechanism(s)
   responsible for quenching star formation in dwarf galaxies on the
   cluster potential, resulting in two distinct evolutionary pathways.
   Firstly, we find a significant increase (at the 3$\sigma$ level) in
   the mean equivalent width of H$\alpha$ emission among star-forming
   dwarf galaxies in the infall regions of the Coma cluster and the
   core of Abell 1367 with respect to the overall supercluster
   population, indicative of the infalling dwarf galaxies undergoing a
   starburst phase. We identify these starburst galaxies as the
   precursors of the post-starburst k+A galaxies. 
   Extending the survey
   of k+A galaxies over the whole supercluster region, we find 11.4\%
   of all dwarf ($z$ mag $\!>\!15$) galaxies in the Coma cluster and 4.8\% in
   the Abell~1367 have post-starburst like spectra, while this
   fraction is just 2.1\% when averaged over the entire supercluster
   region (excluding the clusters). This points to a cluster-specific
   evolutionary process in which infalling dwarf galaxies undergo a
   starburst and subsequent rapid quenching due to their passage
   through the dense ICM. In galaxy groups, the star formation in
   infalling dwarf galaxies is instead slowly quenched due to the
   reduced efficiency of ram-pressure stripping.

   We show that in the central $\sim\, 2\, h^{-1}_{70}$~Mpc of the
   Coma cluster, the ($24\!-\!z$) near/mid infra-red colour of
   galaxies is correlated with their optical $(g\!-\!r)$ colour and
   H$\alpha$ emission, separating all mid-infrared (MIR) detected
   galaxies into two distinct classes of `red' and `blue'.
   By analysing the spatial and velocity distribution of galaxies
   detected at 24\m~in Coma, we find that the (optically) red
   24\m~detected galaxies follow the general distribution of `all' the
   spectroscopic members, but their (optically) blue counterparts
   (i) are almost completely absent in the central $\sim 0.5\,
   h^{-1}_{70}$~Mpc of Coma, and (ii) have a remarkable peak in their
   velocity distribution, corresponding to the mean radial velocity 
   of the galaxy group NGC\,4839,
   suggesting that a significant fraction of the `blue' MIR galaxies
   are currently on their first infall towards the cluster.
 The implications of adopting different SFR tracers for quantifying
   evolutionary trends like the Butcher-Oemler effect are also
   discussed.
 \end{abstract}
 

%======================================================================================
 
 \section{Introduction}

The Coma supercluster is the nearest rich supercluster of
galaxies \citep{cr76,gt78}, consisting of two rich Abell clusters,
separated by $30\,h^{-1}_{70}$~Mpc, but connected by a prominent
filament of galaxies and poorer groups \citep[\eg ][]{font84}, which
is part of the supercluster identified as the ``Great Wall'' in the first major
redshift survey of galaxies in the nearby Universe \citep{gh89}. At a
distance of $\sim 100\,h^{-1}_{70}$~Mpc, it affords a closer look at
the properties of individual galaxies (1~kpc\,$\simeq$\,2.1 arcsec), 
but its large angular scale on the sky presents observational
challenges for the narrow fields of view of most instruments.

It is interesting to note that even though they are the two most
significant structures in the supercluster, the two Abell clusters are
remarkably different in every respect.  The optical luminosity
function of galaxies in the Coma cluster, for instance, has a much
steeper faint-end slope than that of its neighbour
Abell~1367 \citep{paramo03}. Early studies
\citep[e.g.][]{dress80} revealed that the fraction of spiral galaxies
in the Coma cluster is anomalously low compared to 
Abell~1367. In addition, galaxies in the core of the Coma cluster were
identified to be unusually deficient in neutral hydrogen
\citep[e.g.][]{sj78,gh85,bern94}, ionized hydrogen \citep{sr97}
and  molecular gas
(H$_2$/CO) \citep{boselli97a}.
\citet{bothun84} found a strong HI gradient in Coma galaxies-
 many of the inner, HI poor spirals being quite blue, suggesting that some
 gas removal process has acted quite recently, while Abell~1367 was found to
 be a mixture of HI poor and rich galaxies. 

 The availability of more detailed analyses of star formation, based
 on multi-wavelength data, and of spectral indices sensitive to
 stellar ages, has made this field more interesting.  Estimates of
 star formation activity from mid-infrared (IR) \citep{boselli97b} and
 Galaxy Evolution Explorer ({\em GALEX}) ultraviolet \citep{cortese08}
 photometry show that star formation in Coma on the whole is
 substantially suppressed compared to that in the field, while
 Abell~1367 has an abundance of bright star-forming galaxies
 \citep{cortese08}.

In a hierarchical model of the formation of structures in the
Universe, it is not surprising that a rich cluster, with a relaxed
appearance in its X-ray image, such as Coma, would represent the end
product of the gradual assimilation of several galaxy groups over
time.  In spite of the claim of \citet{ds88} that the Coma cluster
does not have significant substructure, it has subsequently been shown
to have at least three major subclusters in the optical
\citep[\eg][]{west98} and X-ray maps \citep[\eg][]{dm93,adami05}.  At
the other end of the supercluster, Abell~1367 is also found to be
elongated with three major subclusters, along the axis of the
filament, with a population of star-forming galaxies infalling into
the SE cloud and possibly the other two as well \citep{cortese04}.
This has prompted a multitude of studies attempting to link the
dynamical history of the supercluster with the properties of
individual galaxies to investigate both the process of building
clusters, as well as the effect of large-scale structure on the star
formation history of galaxies \citep[\eg][]{chris06}.

 By employing the spectral analysis of galaxies in Coma, together with
 the X-ray map of \citet{neumann03}, \citet{poggianti04} found that the
 post-starburst (k+A) galaxies, in which star formation has been
 quenched within the last 1-1.5 Gyr, are associated with the X-ray
 excess attributed to the substructure in Coma. However, this work was
 limited to 3 fields in the $\sim\!2$~Mpc region surrounding the
 centre of Coma.
  
 In addition to star formation indices measured from optical spectra,
 the availability of {\em Spitzer} MIPS mid-IR observations has been
 recently utilised by several authors to characterise the star formation
 properties of the obscured component in galaxies in clusters
 and groups \citep[e.g.][]{saintonge08,chris09,wolf09,bai10}, since
 the 24\m\ flux can be used as a representation of the
 dust-reprocessed overall IR flux. 

 Hereafter, we will refer to the pair of clusters Coma and Abell~1367,
 along with the associated filament of galaxies as the Coma
 Supercluster.  In this paper, we use the {\em Spitzer} MIPS 24\m\
 observations, wherever available, along with the broadband colours
 and spectral star formation indicators from the Sloan digital sky
 survey (SDSS).  We adopt the distance modulus of the Coma
 Supercluster to be $m\!-\!M\!=\!35.0$, and use cosmological
 parameters $\Omega_{\Lambda}\!=\!0.70, \Omega_{M}\!=\!0.30$, and
 $H_0=70$ \kmsmpc\ for calculating the magnitudes and distances. We
 note that at the redshift of Coma (z\,$=\!0.023$), our results are
 independent of the choice of cosmology.  In the next section, we
 present our data and reduction methodology, and summarise our main
 results in \S\ref{coma-sf} and \S\ref{results}. We discuss their
 implications, and compare the properties of galaxies in various parts
 of the supercluster in \S\ref{discussion}, summarising in
 \S\ref{conclusions}.

 \section{Observational Data}
 \label{data}
 
 We base our work on the photometric and spectroscopic data acquired by the
 SDSS Data Release~7 \citep[][(DR7)]{adelman06}, which for the first
 time covers the entire region of the Coma Supercluster presented in
 this paper. Over a smaller area, we also use archival 24\m~mid IR
 (MIR) images, from the Multi-band Imaging Photometer (MIPS)
 instrument aboard {\it{Spitzer}} \citep{rieke}, available over a significant
 fraction of the two main clusters.
 
 \subsection{Optical data}
 
 We select the galaxies belonging to the Coma supercluster from the
 SDSS spectroscopic catalogue only, requiring the member galaxies to
 be within $170.0\!\le$RA$\le\!200.0$ deg and
 $17.0\!\le$Dec$\le\!33.0$ deg on the sky, and with a radial
 velocity within 2,000 km~s$^{-1}$ of the mean redshift of the Coma
 cluster (6,973 km~s$^{-1}$) or the Abell 1367 cluster (6,495
 km~s$^{-1}$) respectively \citep{rines03}.  All our galaxies are
 brighter than SDSS magnitude $r\!=\!17.77$ ($\sim\, M^{*}\!+\!4.7$ for
 Coma), which is the completeness limit of the SDSS spectroscopic
 galaxy catalogue.

 \subsection{MIPS 24\m\ data}
 
\begin{figure*}
 \centering{
 {\rotatebox{270}{\epsfig{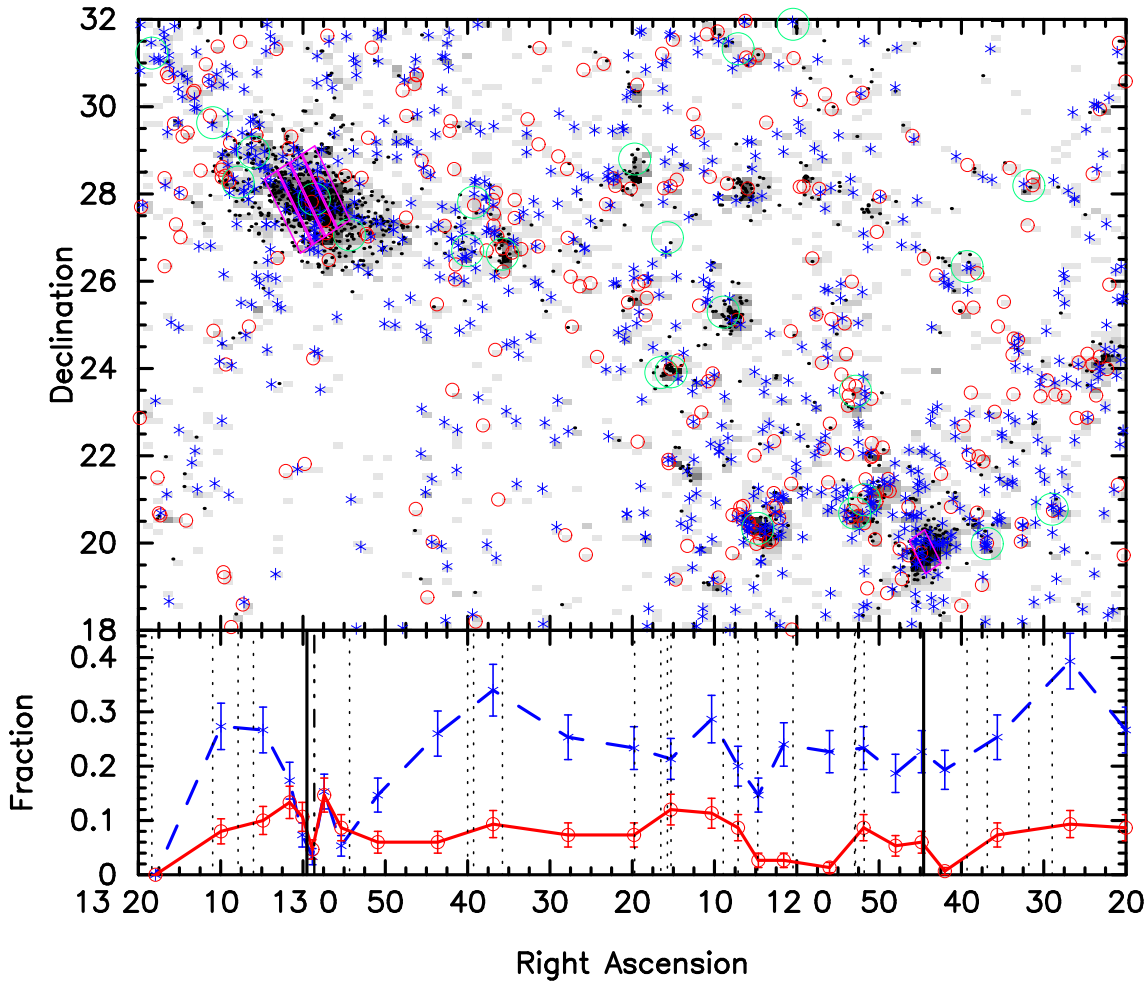}}}}
\caption{{\bf{Top panel:}} The surface density of galaxies in the Coma
  supercluster is shown in {\it grey}. The positions of passive
  galaxies ({\it{black dots}}), AGN host galaxies ({\it{open red
      circles}}) and starburst galaxies {\it{(blue stars)}};
  EW(H$_{\alpha})\geq 25$\AA$\sim$ log SSFR$\sim -10$yr$^{-1}$) are
  indicated. The {\it{big green circles}} are the groups in the
  region, from the NASA Extragalactic Database.  The rectangular pink
  regions show the Spitzer MIPS fields of view for the two
  clusters. {\bf{Bottom panel:}} The {\it{solid red}} line and the
  {\it{dashed blue}} line show the fraction of AGN and of the
  starburst galaxies, among all the galaxies shown in the upper
  panel. The bins are chosen to have 150 galaxies in each of them.
  The solid vertical lines represent the centres of the Coma cluster
  and Abell~1367 respectively, while the {\it dot-dashed line} shows
  the RA position of the NGC\,4839 group (see text).  All the dotted
  lines represent the positions of groups, shown  %as {\it{green circles}}
  in the upper panel. }
 \label{scl}
 \end{figure*}

 For the mid-infrared study of Coma and Abell 1367, we use archival
 24\m~{\em Spitzer} MIPS data covering $2{\times}2\,{\rm deg}^{2}$ in
 the case of Coma (PID: 83, PI G.~Rieke) and
 $30^\prime{\times}30^{\prime}$ for Abell 1367 (PID: 25, PI
 G.~Fazio). The Coma 24\m\ dataset consists of four contiguous mosaics
 (see Fig.~\ref{scl}) obtained in medium scan mode, with scan leg
 spacing equal to the half array width, producing homogeneous
 coverage over the mosaic, with an effective exposure time per pixel
 of 88\,sec. \citet{bai}, who used this data to determine the
 24\m~luminosity function of Coma, estimate the 80\% completeness limit
 to be 0.33\,mJy, corresponding to a star formation rate (SFR) of
 0.02\,M$_{\odot}\,{\rm yr}^{-1}$ at the redshift of Coma.  The Abell
 1367 24\m\ dataset consists of a single mosaic obtained in medium scan
 mode, with scan leg spacing equal to the full array width, producing
 an effective exposure time per pixel of 40\,sec.

The {\sc SExtractor} package \citep{bertin} was used to automatically detect
 sources, and obtain photometric parameters. The images were first
 background-subtracted, and then filtered with Gaussian functions,
 with full width half maximum (FWHM) matching the 24\m\ point-spread
 function (PSF). Aperture photometry was obtained for all objects
 having 8 contiguous pixels above the 1$\sigma$ rms background noise
 level. Following SWIRE, we measured the fluxes within circular
 apertures of diameter 21, 30, 60, 90 and 120$^{\prime\prime}$.  For
 the vast majority of sources, the 6$^{\prime\prime}$ FWHM
 point-spread function of MIPS leaves the object unresolved at 24\m,
 and hence we estimate the total 24\m\ flux of these objects from the
 flux contained within the 21$^{\prime\prime}$ aperture, corrected by
 a factor 1.29.

 However, at the redshift of the Coma and Abell 1367 clusters, many of
 the brighter galaxies (spirals in particular) are larger than our
 nominal 21$^{\prime\prime}$ diameter aperture, such that a
 significant fraction of their 24\m\ flux is missed. For these
 galaxies, we use instead one of the larger apertures, matched to
 contain the optical flux as parametrized by R$_{90z}$ (the radius
 which contains 90\% of the $z-$\,band flux) as measured from the SDSS
 image. Finally, for some of the bright cluster galaxies (9 in total),
 the value of R$_{90z}$ is under-estimated. To correct for this, SM
 eyeballed the 24\m\ images of these objects, and estimated an
 appropriate radius, in each case, for extracting the `total' 24\m\
 flux. In the case of the brighter spiral galaxies in Coma and Abell
 1367, the 24\m\ emission often shows significant structure due to
 spiral arms and star-forming regions, necessitating a high value of
 the deblending parameter, to prevent the galaxy being shredded by
 {\sc SExtractor}. Particular care was required to correctly deblend
 the face-on spiral NGC\,3861 from NGC\,3861B, while keeping NGC\,3861
 intact. We note that for $\sim$\,50\% of our sample the typical
 measurement error in the 24\m\ flux is $<\!0.1$ mJy (which is
 $\lesssim\!10\%$ of the measured flux), and for $>\!95\%$ the error
 is $<\!0.35$ mJy. In total, $\sim$\,90\% of our galaxies have 24\m\
 flux measured with $<\!30\%$ uncertainty.

 \subsection{Matching the SDSS spectroscopic catalogue with the 24\m\
   sources}
 
 To match the SDSS spectroscopic catalogue for the Coma supercluster
 with the MIR 24\m\ data, we choose the nearest optical counterpart
 within 5$^{\prime\prime}$ of each 24\m\ source.
 This accounts for a displacement of no more
 than 2.5~$h^{-1}_{70}$ kpc between the centres. We note that
 $\gtrsim$80\% of the matches are found to have centres within
 2$^{\prime\prime}$ of each other. 16 of our sources above the
 detection threshold were found to have multiple matches within
 the matching radius of 5$^{\prime\prime}$.
 
 Our final dataset comprises of 3,787 galaxies ($r\!\leq\!17.77$) with
 spectroscopic redshifts from SDSS, and radial velocity within
 $\pm$2,000 km~s$^{-1}$ of Coma or Abell 1367, in the $\sim$500 square
 degree region of the sky covered by the Coma supercluster.  Out of
 these, 197 within $\sim$2 $h^{-1}_{70}$~Mpc of the centre of Coma and
 24 within $\sim$0.65 $h^{-1}_{70}$~Mpc of the centre of Abell 1367
 are found to have a 24\m\ counterpart. The probability that a background
 MIPS source lies within the matching radius of 5$^{\prime\prime}$
 by chance, depends only on the density of the 24\m\ sources. For the
 completeness limit (0.33 mJy) adopted here, this density is
 $\sim\!3,000$\,deg$^{-2}$ \citep[see fig. 3 of][]{shupe08}. This gives
 the required probability as 0.018, implying that our sample of 197
 24\m\ detected galaxies in Coma may have 3--4 interlopers.
 A sample of the catalogue of data for the 24\m\ detected galaxies is
 provided in Table~\ref{tbl:sb-table}.

 \begin{table*}
 \begin{minipage}{\linewidth}
 \centering{
 \caption{Catalogue of Coma galaxies detected at 24$\mu$m 
(This table is available in full online.)}
\begin{tabular}{|c|c|c|c|c|c|c|c|}
 \hline
 \hline
 RA       &  Dec     & z      & $g$      & $r$      & $z$ & 24$\mu$m flux & flag\footnote{
 Classification on the basis of BPT diagram,where \\
 0: unclassified \\
 1: star-forming \\
 2: AGN \\
 3: AGN according to \citet{miller03} criteria (see text) \\} \\
 (J2000)  & (J2000)  &        & mag      & mag      & mag & (mJy)         &   \\
 \hline
 194.4574 &  28.6243 & 0.0250 & 17.80 & 17.35 & 17.24 &   0.31 & 1 \\ 
 195.1548 &  28.6641 & 0.0236 & 17.47 & 16.89 & 16.49 &   1.25 & 1 \\ 
 194.5385 &  28.7086 & 0.0255 & 15.21 & 14.56 & 14.08 &  36.02 & 1 \\ 
 194.9172 &  28.6308 & 0.0179 & 15.67 & 15.47 & 15.30 &   7.90 & 1 \\ 
 194.3645 &  28.4397 & 0.0260 & 17.16 & 16.57 & 16.19 &   0.75 & 1 \\ 
 194.5628 &  28.5218 & 0.0231 & 17.00 & 16.56 & 16.32 &   1.84 & 1 \\ 
 194.3912 &  28.4823 & 0.0209 & 14.35 & 13.54 & 13.00 &   1.86 & 3 \\ 
 194.4747 &  28.4998 & 0.0244 & 15.94 & 15.18 & 14.56 &   0.04 & 0 \\ 
 195.2170 &  28.3661 & 0.0255 & 14.51 & 13.73 & 13.10 &   1.39 & 0 \\ 
 195.3123 &  28.5217 & 0.0281 & 17.48 & 17.11 & 16.98 &   1.06 & 1 \\ 
 \hline
 \label{tbl:sb-table}
 \end{tabular}}
 \end{minipage}
 \end{table*}

 \section{Star formation and AGN across the Coma supercluster}
 \label{coma-sf}

 Located $\sim$100~Mpc from us, the Coma supercluster
 offers a unique opportunity for investigating the effect on individual
 galaxies of the hierarchical formation of structures in the Universe,
 as galaxies progress through various environments towards the cores
 of the clusters.

 In order to give a general overview of the optical star formation
 properties of galaxies along the entire supercluster, in
 Fig.~\ref{scl} we plot the positions of all the galaxies from the
 SDSS spectroscopic catalogue found in our redshift range, and the
 major galaxy groups obtained from the NASA Extragalactic Database
 (NED, http://nedwww.ipac.caltech.edu/). In order to avoid multiple
 detections, we consider the groups only from the
 NGS, WBL, USGC and HCG catalogues,
 and even amongst these, we identify duplication by merging groups
 closer than 1$^{\prime}$ and within $\pm$100 km~s$^{-1}$
 of each other. 

 The positions of AGN host galaxies, identified using the BPT diagram
 \citep[][also see \S\ref{sf}]{bpt81}, and non-AGN starburst galaxies
 (star-forming on the BPT diagram, and having
 EW(H$_\alpha$)$\geq$25\AA), are indicated in Fig.~\ref{scl}. 
 We also show the limited regions at the cores of the Coma and Abell 1367
  clusters, over which the 24\m\ data are available, and utilised 
  in this paper. 
 Plotted in the bottom panel of the same figure, is the distribution
 of the AGN and starburst galaxies, as fractions of the 150 galaxies
 contributing to each bin. It is worth mentioning that although these
 distributions are a fair representation of the data, the binning has
 resulted in a spurious feature at the position of Abell 1367.  As is
 already well known in the literature, the central region of Abell 1367 is
 sparsely populated, the majority of galaxies in the core being 
 late-type. In this figure, it appears that the fraction of
 starburst galaxies does not decline in the core of Abell 1367, but if
 the data is re-binned so as to have equally-spaced bins, this would not
 be the case. Having said that, the decline in the fraction of
 starburst galaxies at the centre of Abell~1367 is much shallower than
 that seen in Coma.
 
 \begin{figure*}
 \centering{{\epsfig{file=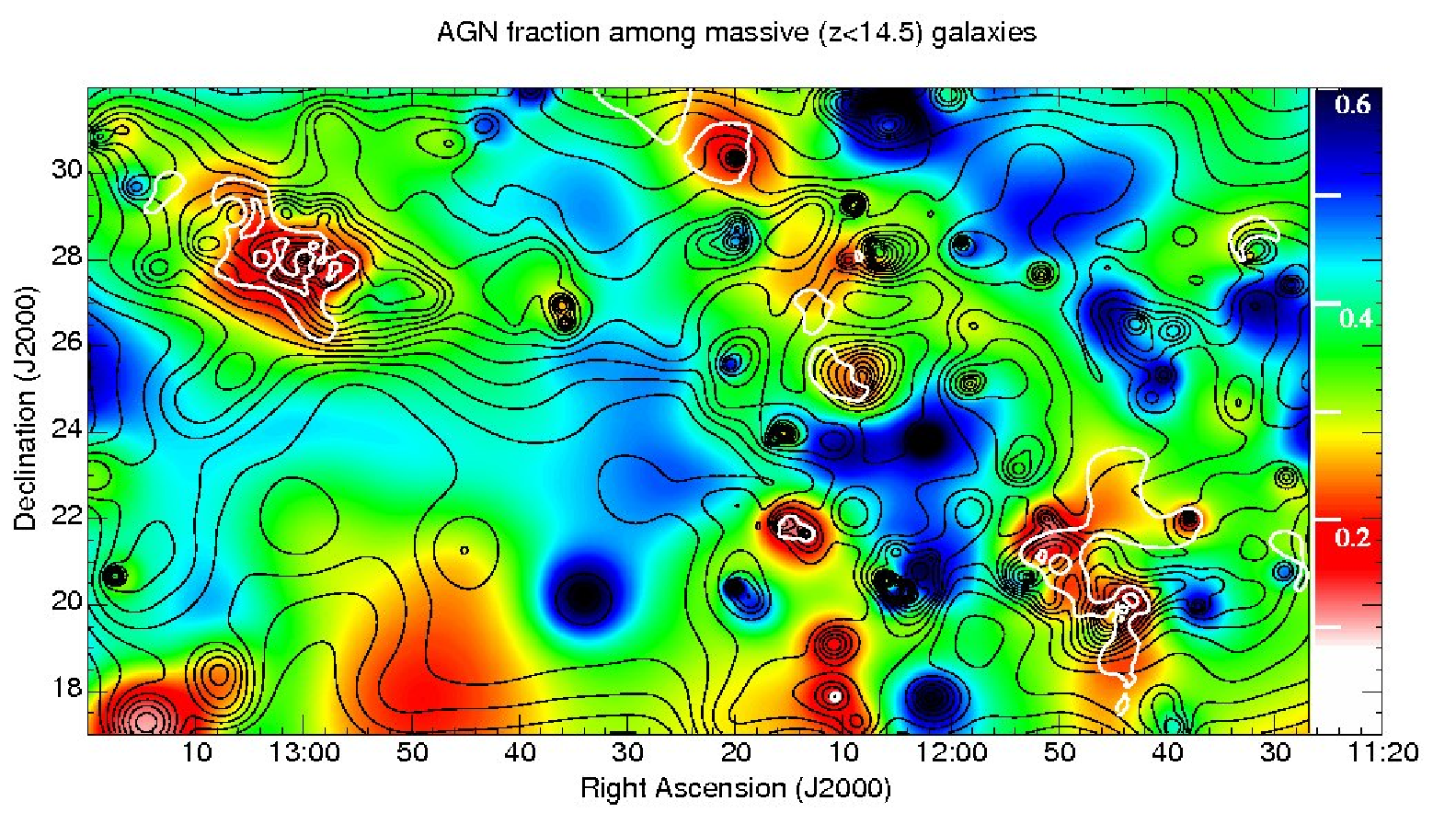,width=17cm}}}
 \caption{The local AGN fraction ($f_{AGN}$) among massive galaxies
   ($z\!<\!14.5$), as a function of spatial position across the Coma
   supercluster. The map is colour-coded with $f_{AGN}$.
   The {\it{black contours}} indicate the
   local luminosity-weighted ($z$-band) galaxy density across the
   supercluster.  The {\it{white contours}} indicate regions with AGN
   fraction $f_{AGN}=$ 1 and 2-$\sigma$ below the mean
   value over the supercluster. }
  \label{agn_frac}
 \end{figure*}

 In general, the optically-identified AGN seem to be more or less
 uniformly distributed throughout the supercluster (Figs.~\ref{scl}
 and \ref{agn_frac}), except for a sharp decline in the cores of both
 the clusters. This latter result is counter-intuitive, because AGN
 hosts are known to be early-type massive galaxies which mostly reside
 in dense environments. Hence, this may indicate that the optical
 emission from the AGN present in this region is obscured and/or
 detectable at other wavelengths (radio/X-ray).
 The latter is beyond the scope of this work, but our
 MIPS 24\m\ data supports the former possibility. As seen in
 Fig.~\ref{radius}, not only does the relative fraction of the red
 24\m\ galaxies marginally increase towards the centre, around 50\% of
 them are found within 0.5~$h^{-1}_{70}$~Mpc of the centre of the Coma
 cluster, indicating that optical emission from several AGN hosts may
 be obscured. Further evidence in support of this argument may be
 drawn from Fig.~\ref{bpt} where a large fraction of galaxies detected
 at 24\m\ do not show [OIII] and/or H$_\beta$ in emission, but their
 H$_\alpha$/[NII] flux ratios are consistent with the presence of an
 AGN \citep{miller03} on the BPT diagram \citep{bpt81}.

 We further investigate the environmental trends for AGN and star
 formation activity within the Coma supercluster, by estimating the
 local fraction of star-forming galaxies ($f_{SF}$) and AGN
 ($f_{AGN}$) over the entire supercluster, both for massive galaxies
 \citep[$z\!<\!14.5$; $M_{z}{<}M^{*}+1.8$, assuming
 $M_{z}^{*}{=}-22.32$ from][]{blanton01} and the dwarf galaxy
 population ($z>15$; $M_{z}>M^{*}+2.3$). Following \citet{chris07}, we
 define the local projected density of galaxies ($\rho(\mathbf{x})$)
 using a variant of the adaptive kernel estimator \citep{silverman86,
   pisani93}, where each galaxy $i$ is represented by an adaptive
 Gaussian kernel $\kappa_{i}(\mathbf{x})$. This is different from the
 algorithm of \citet{silverman86}, which, in its previous applications
 to astronomical data, requires the kernel width $\sigma_{i}$ to be
 iteratively set to be proportional to $\rho_{i}^{-1/2}$. In this
 work, we fix the transverse width $\sigma_{i}$ to be proportional to
 $D_3$, where $D_3$ is the distance to the third nearest neighbour
 within 500\,km\,s$^{-1}$.

 The dominant factors governing the star formation properties of a
 galaxy are the mass of its DM halo, and whether it is the central
 or a satellite galaxy in a halo \citep[\eg][]{kauffmann04,
 yang05, blanton06}.  We adopt the above method, and choose the
 dimensions of the kernel, keeping this in mind.  In the case of
 galaxies within groups or clusters, the local environment is measured
 on the scale of their host halo (0.1--1 Mpc), while for galaxies in
 field regions the local density is estimated by smoothing over their
 5--10 nearest neighbours or on scales of 1--5\,Mpc \citep[for details, see][]{chris07}.

\begin{figure*}
 \centering{{\epsfig{file=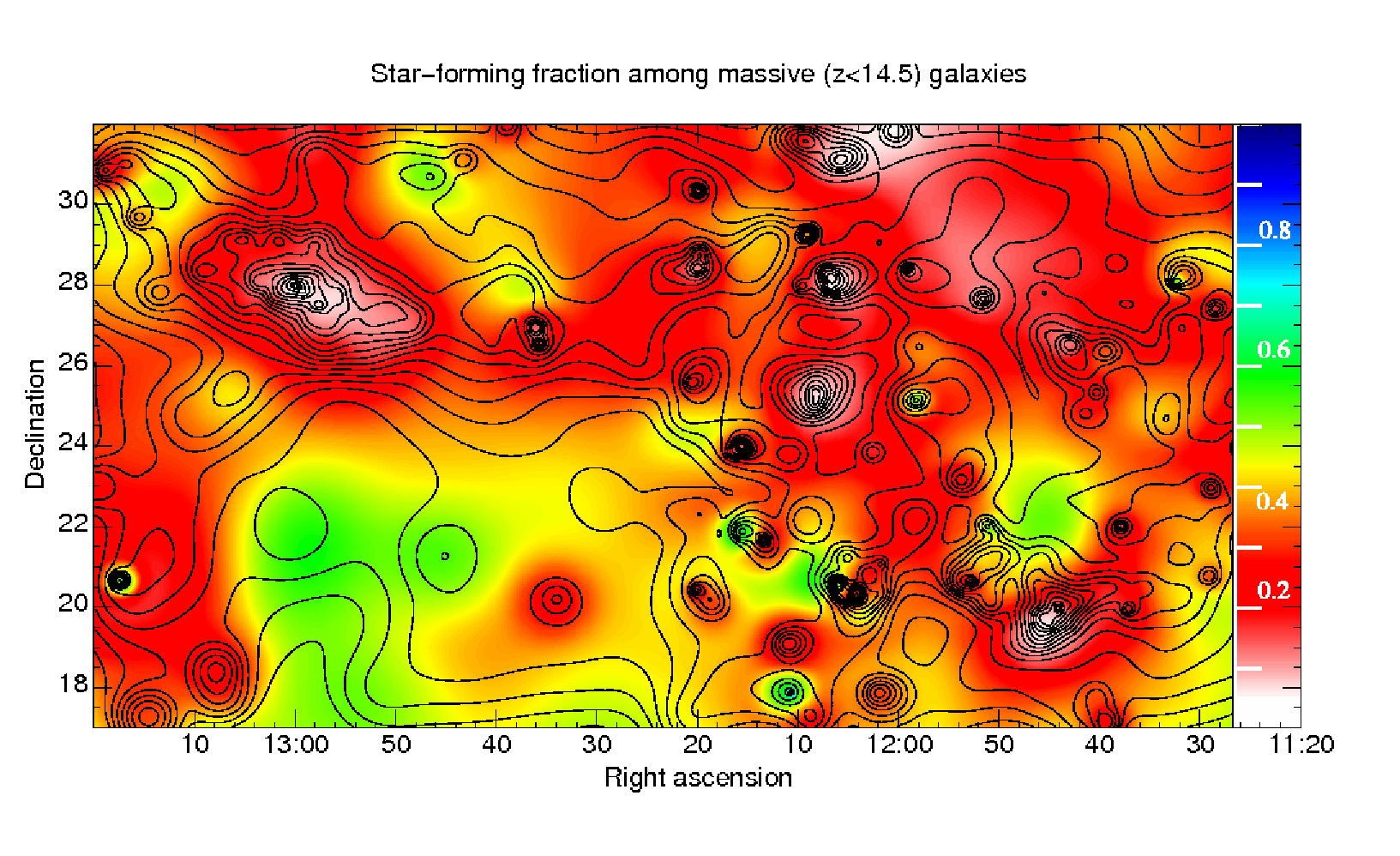,width=17cm}}}
 \centering{{\epsfig{file=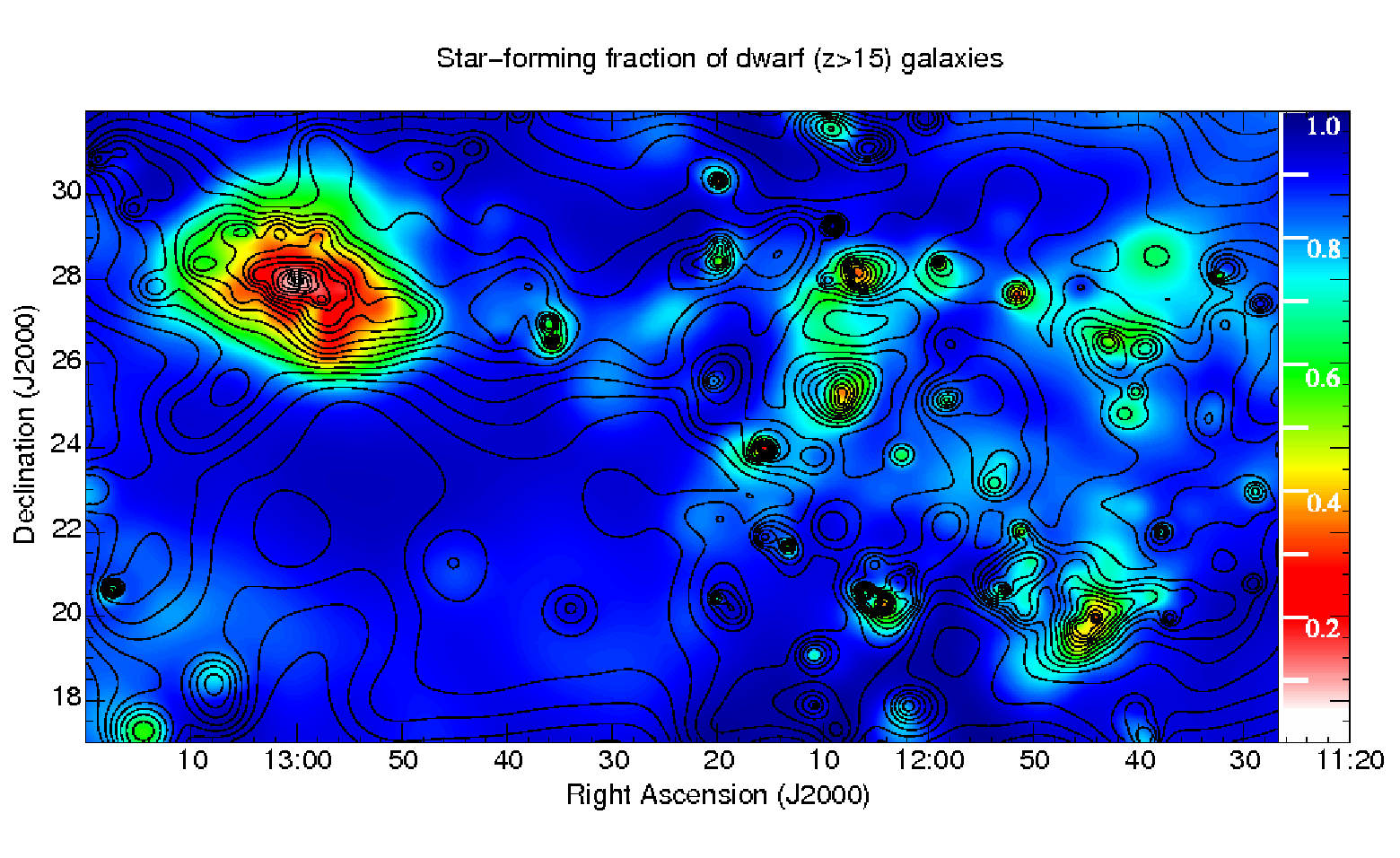,width=17cm}}}
 \caption{The local fraction of star-forming galaxies ($f_{SF}$) as a
   function of spatial position across the Coma supercluster, for
   massive galaxies ($z\!<\!14.5$; {\em top panel}) and dwarf galaxies
   ($z\!>\!15$; {\em lower panel}). The colours indicate $f_{SF}$, and
   the same colour scale is used in
   both the panels. Overlaid are black contours indicating the
   $z$-band luminosity-weighted galaxy density across the
   supercluster. This figure shows that in the Coma supercluster, star
   formation in the massive galaxies ($z\!<\!14.5$) seems to be
   suppressed independent of their local environment, while the dwarf
   galaxies ($z\!>\!15$) are star-forming everywhere except in the
   dense environment in the vicinity of rich clusters and
   galaxy groups. 
}
 \label{sf_frac}
 \end{figure*}

\begin{figure*}
 \centering{{\epsfig{file=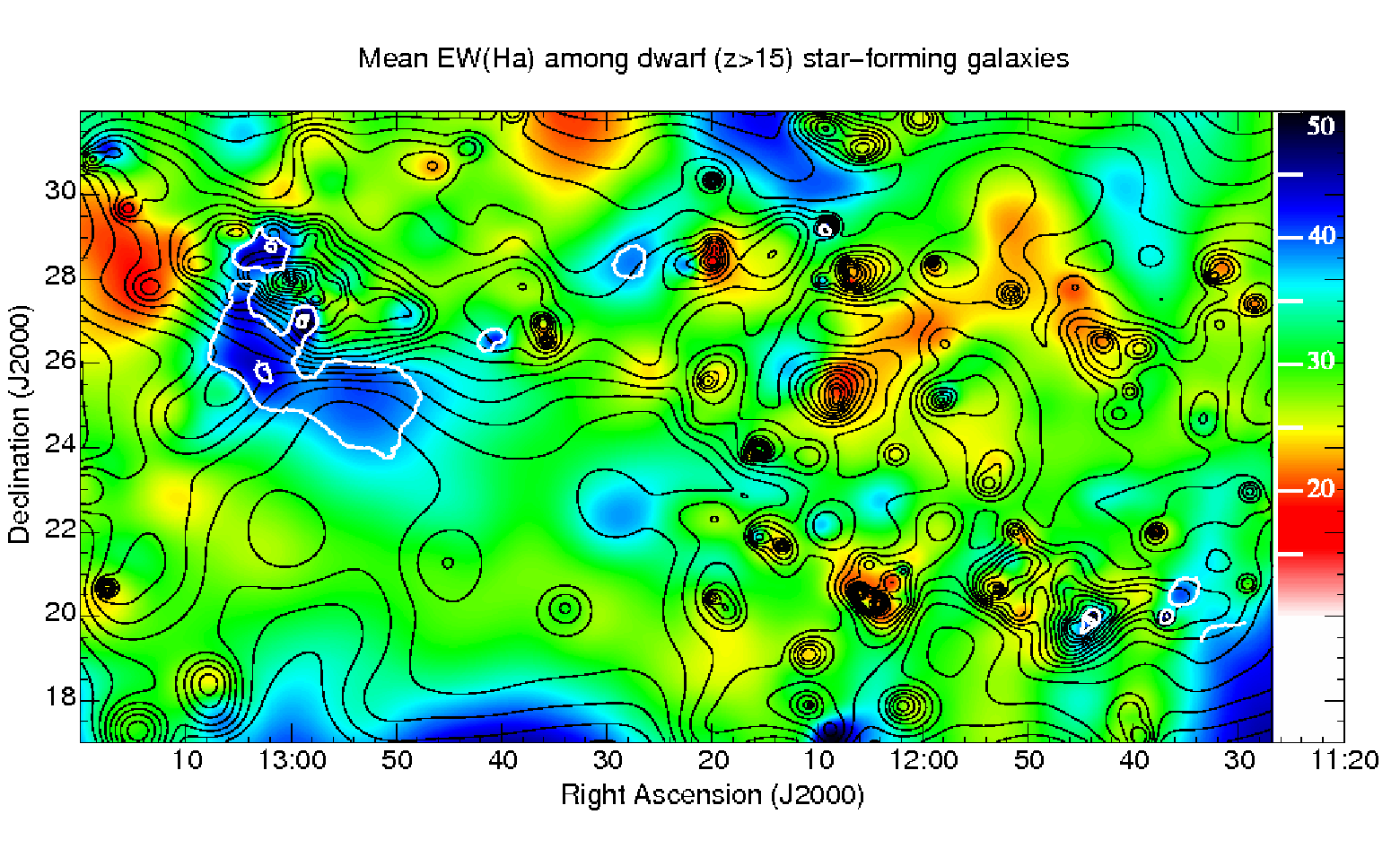,width=15cm}}}
 \caption{The local mean H$_\alpha$ equivalent width of star-forming
   (EW(H$_\alpha)\!>\!2$\AA) dwarf ($z\!>\!15$) galaxies across the
   Coma supercluster, indicated by colour. Overlaid are {\it{black contours}}
   indicating the $z$-band luminosity-weighted galaxy density across
   the supercluster, and {\it{white contours}} showing regions where
   the local observed mean value of the EW(H$_\alpha$) is 2 and
   3-$\sigma$ above that averaged over the whole supercluster. }
 \label{mean_ha}
 \end{figure*}

 We can then map the local galaxy density ($\rho$) and $z$-band
 luminosity density ($j_z$) as
 \mbox{$\rho(\mathbf{x})=\sum_{i}\kappa_{i}(\mathbf {x{-}x}_{i})$} and
 \mbox{$j_{z}(\mathbf{x})=\sum_{i}L_{z,i}\,\kappa_{i}(\mathbf
   {x{-}x}_{i})$}, where $L_{z,i}$ is the $z$-band luminosity of
 galaxy $i$ (see Fig.~\ref{sf_frac}). Following \citet{yang05}, we
 identify local maxima in the $z$-band luminosity density as galaxy
 groups and clusters, whose masses correlate with the total $z$-band
 luminosity associated with the peak.  By comparison with the
 Millennium simulation, we expect all groups with 4 or more members to
 be associated with a local maximum in \mbox{$j_{z}(\mathbf{x})$}
 \citep{chris07}. The local $z$-band luminosity density is shown as
 black contours in Figs.~\ref{agn_frac}--\ref{mean_ha}.

 In Fig.~\ref{agn_frac} we show the spatial variation of $f_{AGN}$
 among massive galaxies ($z\!<\!14.5$) across the supercluster, where
 we calculate $f_{AGN}$ as
 $\sum_{i{\subset}AGN}\kappa(\mathbf{x{-}x}_{i})/\rho(\mathbf{x})$,
 where $i{\subset}AGN$ is the subset of galaxies classified as optical
 AGN, based on the ratios of thermally excited and recombination lines
 used in the BPT diagram.  We do not consider the lower mass galaxies,
 as the fraction of AGN declines rapidly to zero below
 $M_z\!>\!M^{*}\!+\!2$;

 For comparison, Fig.~\ref{sf_frac} shows the distribution of
 $f_{SF}(\mathbf{x})$ as
 $\sum_{i{\subset}SF}\kappa(\mathbf{x{-}x}_{i})/\rho(\mathbf{x})$,
 where $i{\subset}SF$ is the subset of galaxies classified as
 star-forming according to the line-ratios used in the BPT
 diagram. Fig.~\ref{mean_ha} shows a map of the local mean equivalent
 width (EW) of H$\alpha$ of star-forming galaxies as
 $\langle{EW(H\alpha)}(\mathbf{x})\rangle=\sum_{i{\subset}SF}{\rm
   EW(H}\alpha)\,\kappa(\mathbf{x{-}x}_{i})/\sum_{i{\subset}SF}\kappa(\mathbf{x{-}x}_{i})$,
 plotted in a manner analogous to that used by \citet{chris06} for
 galaxy colours in the Shapley supercluster.

 In Fig.~\ref{agn_frac} we see that the AGN fraction $f_{AGN}$,
 among massive galaxies ($z\!<\!14.5$) declines at
 the centre of the Coma and Abell 1367 clusters, while elsewhere, it is
 uniformly distributed. 
This appears to contradict the results of
 \citet{chris07}, who found the AGN fraction to be independent of
 environment, and to be a monotonically increasing function of stellar
 mass. However, the volume studied by \citet{chris07} within SDSS DR4
 did not cover clusters as rich as Coma or Abell 1367. This may
 indicate that only the very dense environments affect the optical AGN
 activity of galaxies. The relation between $f_{AGN}$ and local
 environment is unclear in the intermediate density group
 environment. While the $f_{AGN}$ values decline in the centre of some
 groups, in others they exceed the mean value for the field. Although
 there is a slight indication that the groups in which the $f_{AGN}$
 appears to be declining, mostly lie on the filament connecting the Coma
 and Abell 1367, or in the vicinity of the clusters themselves, while
 the groups with higher values of $f_{AGN}$ lie in underdense regions.
 However, it is not possible to draw any firm statistical inferences
 from this observation.

 Whether the AGN fraction $f_{AGN}$ varies with environment or not
 depends on how the activity of the galactic nuclei is defined.
 \citet{miller03} do not find any correlation between galaxy density
 and f$_{AGN}$ in a large sample of galaxies ($M_r\!=\!-20;
 0.05\!<\!\rm z\!<\!0.095$), where the AGN are defined in terms of
 optical emission line ratios characterised by the BPT diagram.  Other
 studies find higher incidence of optical AGN in groups and clusters
 \citep{arnold09}.  If the AGN were selected according to their radio,
 mid-IR or X-ray properties, they would be found in differen hosts:
 radio AGN in early-type galaxies, IR AGN in bluer galaxies, and X-ray
 AGN in galaxies of intermediate colour \citep{hickox09}. The
 environment dependence of AGN activity is thus largely linked to the
 distribution of the hosts.  The incidence of X-ray AGN is higher in
 galaxy groups than in galaxy clusters \citep{martini06,shen07}. The
 fraction of radio-loud AGN is the same in the brightest galaxies of
 groups and clusters and in the field, but higher in non-central
 galaxies \citep{best05}.  \citet{wdp10} show that the mass function
 of black holes is independent of environment, and the variation in
 the distribution of optical, radio and X-ray AGN can be understood in
 terms of the accretion processes that lead to the manifestation of
 the AGN in the various ranges of electromagnetic radiation.

 We measured the significance of the spatial variations seen in
 $f_{AGN}$ across the supercluster by performing Monte Carlo
 simulations, in which we made repeated maps of $f_{AGN}$ after
 randomly assigning the positions of the AGN to the bright galaxies
 (\ie\ testing the null hypothesis that $f_{AGN}$ is constant across
 the supercluster), and measuring the fraction of maps in which a
 given local value of $f_{AGN}$ was obtained within one of the Monte
 Carlo simulations. The results of these simulations are represented
 by the white contours, which indicate regions that have $f_{AGN}=$
 1 and 2-$\sigma$ below the mean value across the
 supercluster. This confirms that the decline in $f_{AGN}$ seen
 towards the core of the Coma cluster is significant at the 3$\sigma$
 level, while that seen in Abell 1367 is significant at the 2$\sigma$
 level.

 In Fig.~\ref{sf_frac} we show the general correlation between the
 fraction of star-forming galaxies ($f_{SF}$) and environment for
 massive galaxies ($z\!<\!14.5$; top panel) and the dwarf galaxies
 ($z\!>\!15$; bottom panel). For the massive galaxies we find an
 almost uniform $f_{SF}\!<\!0.5$ in all the environments. For dwarf
 galaxies, we see a much stronger SF-density relation, with $f_{SF}$
 rising rapidly from the $f_{SF}\!<\!0$.1--0.4 seen in the cluster
 cores, to $f_{SF}\!>\!0.95$ in the field. This result is in agreement
 with literature, where, using colour or SFR indicators,
 it has been shown that dwarf galaxies exhibit much stronger
 radial trends with environment than their massive counterparts
 \citep[][among others]{gray04,tanaka,smith06,chris07}.
 The origin of the SF-density relation for the
 dwarf galaxies can be attributed to the fact that (i) the star
 formation in dwarf galaxies can easily be quenched by the tidal
 impact of a massive neighbour and/or the ICM of the cluster
 \citep[\eg][]{larson}, and (ii) unlike their massive counterparts,
 dwarfs (of the luminosities considered here) do not become passive
 by internal mechanisms such as gas consumption
 through star formation, merging etc.. However, the effects of supernovae
 wind blowouts and feedback become significant for relatively faint
 ($M^*\!\lesssim\!10^7$ M$_\odot$) dwarfs \citep[\eg][]{maclow99,marcolini06}. 
 In this work we do not find any evidence for the quenching of star
 formation in galaxies on the filament(s), but only in the cores of
 the embedded galaxy groups.

 The H$_\alpha$ emission traces the current SFR of a galaxy, while the
 continuum flux under this line can be used as an indicator of its
 past SFR, making the globally averaged EW(H$_\alpha$) from a galaxy
 an effective indicator of the birthrate parameter ($b\!\equiv$
 current SFR normalized by the SFR averaged over the lifetime of a
 galaxy). Recently, \citet{lee09} have shown that
 EW(H$_\alpha$)\,$\sim\!40$\AA~corresponds to $b\!=\!1$.  In
 Fig.~\ref{mean_ha} we show the variation in the mean EW(H$_\alpha$)
 as a function of environment for the star-forming dwarf galaxies.
 This is potentially a very powerful technique for dissociating the
 intrinsically active star formation history (SFH) of low-mass
 galaxies from a starburst caused by the impact of local environment
 \citep[see][for instance]{chris07}. 

If galaxies are slowly quenched by
 interactions with their environment \citep{balogh04a}, a decline in
 the mean EW(H$_\alpha$) is expected in the regions of denser
 environment. On the other hand, if star formation is
 triggered due to the impact of environment, the
 mean EW(H$_\alpha$) should increase.  Unlike the slow quenching of
 dwarfs in the dense environments \citep{tanaka,chris07}, here, in the
 Coma supercluster, we find that the dwarf galaxies follow different
 evolutionary paths in the groups and in the denser cluster
 environments. In galaxy groups, a dwarf galaxy is slowly quenched via
 interactions with the other group members and/or the tidal field of
 the group, while in the clusters, an infalling dwarf experiences a
 starburst in the intermediate density environment at the cluster
 periphery \citep{porter08,mrp10}, and is then rapidly
 quenched via cluster-related environmental mechanisms, such as
 ram-pressure stripping.

 In analogy to Fig.~\ref{agn_frac}, we measure the significance of the
 spatial variations seen in the mean EW(H$_\alpha$) via Monte Carlo
 simulations, this time by making repeated maps after randomly
 swapping the values of EW(H$_\alpha$) among the dwarf star-forming
 galaxies. In doing this, we seek to test the null hypothesis that the
 EW(H$_\alpha$) is independent of spatial position among star-forming
 galaxies. In Fig.~\ref{mean_ha}, we show, as overlaid white contours,
 the regions in which the local observed mean value of the
 EW(H$_\alpha$) is 2 and 3$\sigma$ above that averaged over the whole
 supercluster. This confirms that the excess star formation seen in
 the infall regions of Coma, and towards the core of Abell~1367, is
 indeed significant at $>$\,3$\sigma$ level.  In quantitative terms,
 this excess in the infall regions of the Coma cluster is due to a
 population of $\sim$\,30 dwarf starburst galaxies ($z\!>\!15$,
 EW(H$_\alpha)\!>\!40$\AA) located within or along the white
 2-$\sigma$ contours in Fig.~\ref{mean_ha}. The analogous excess in
 the core of Abell~1367 can be ascribed to $\sim$\,10 starburst dwarf
 galaxies.

 \section{Optical and mid-IR analysis of Coma and Abell~1367 galaxies}
 \label{results}

 The infrared (IR) emission from normal galaxies around
 $\lambda{=}24\mu$m is dominated by hot dust in the H{\sc ii} regions
 of massive stars, in addition to (usually) minor contribution from
 asymptotic giant branch (AGB) stars and the general interstellar
 medium (ISM).  This recycled emission, together with the optical
 emission, can thus provide a good estimate of the total starlight of
 a galaxy and be used to better constrain the current and past star
 formation rate (SFR) of a galaxy \citep[][among
 others]{calzetti,kennicutt09,rieke09}.  While IR astronomers usually
 study late-type galaxies, early-type galaxies have received sporadic
 attention only in the context of the warm dust component detected in
 a few of them \citep[\eg][]{knapp}.

 In this paper, we combine the optical photometric and spectroscopic
 data taken by the SDSS DR7, with the 24\m\ {\it{Spitzer}}/MIPS data,
 which is sensitive to processed optical emission from stars, to
 study the star formation activity of galaxies residing in the denser
 regions of the Coma supercluster.

 \subsection{Optical and MIR colours}
 \label{colours}
 
 \begin{figure}
 \centering{
 {\rotatebox{0}{\epsfig{file=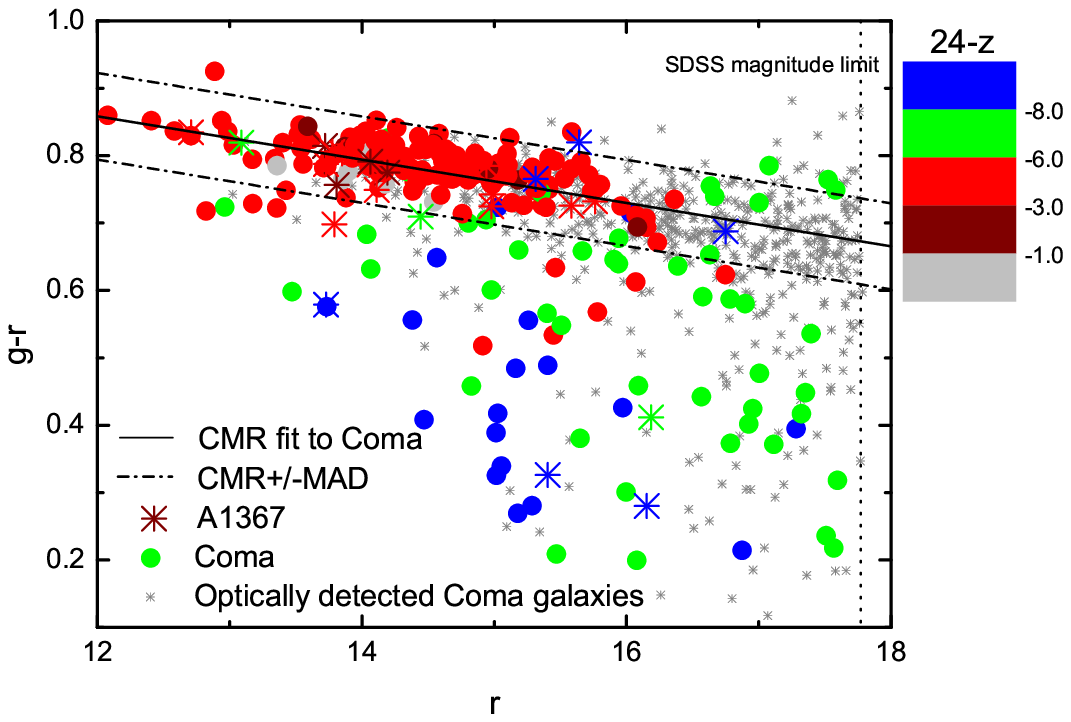,width=10cm}}}}
\caption{The $(g\!-\!r)$ vs $r$ colour-magnitude relation for the
  24\m\ bright galaxies detected in the Coma cluster ({\it{circles}})
  and Abell~1367 ({\it{big stars}}). The solid line shows the
  colour-magnitude relation (CMR) fitted to the Coma data points,
  while the dot-dashed lines mark the mean absolute deviation (MAD)
  boundaries on either side.  We adopt the lower MAD boundary in
  $(g\!-\!r)$ for segregating the red sequence from the blue
  cloud. Note that we do not use any upper bound for the red sequence
  because fitting is done using spectroscopically confirmed cluster
  members only. The symbols are colour-coded according to the
  $(24\!-\!z)$ colour of the MIR bright sources (Fig.~\ref{ssf}). All
  the other Coma galaxies are shown as {\it{grey stars}}. Note that
  $(24\!-\!z)$\,=\,-6 mag separates  the red sequence galaxies from
  the blue ones in a $z\!-\!(24\!-\!z)$ plane (Fig.~\ref{ssf}). }
 \label{cmr}
 \end{figure}

 \begin{figure}
 \centering{ 
{\rotatebox{270}{\epsfig{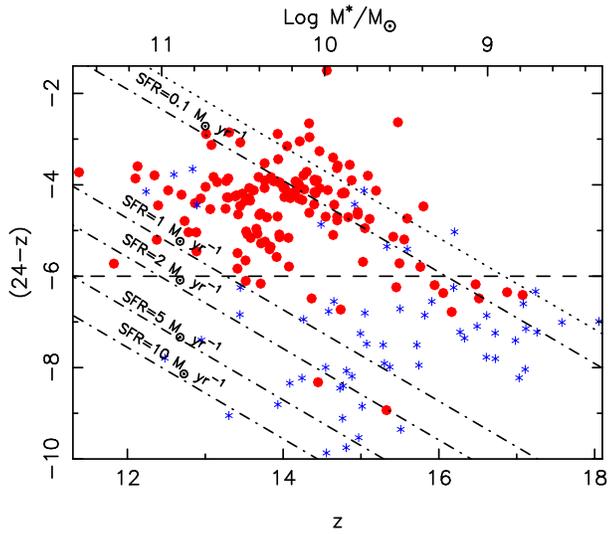}}}}
 \caption{The $(24\!-\!z)$ colour of all galaxies in the Coma
   supercluster sample, divided into blue cloud ({\it blue stars})
   and red sequence ({\it red}) (Fig.~\ref{cmr}) galaxies, are plotted
   against their $z$-band magnitudes. The dotted line shows the 80\%
   completeness limit of the $24\mu$m data. The constant SFR lines
   are drawn using the conversion factor given by
   \citet{calzetti}. The horizontal dashed line at $(24\!-\! z)\!=\!-6$
   mag is our empirically chosen criterion for selecting star-forming
   galaxies. The top axis shows the stellar mass of galaxies estimated
   using a relation from \citet{bell03}. }
 \label{ssf} 
 \end{figure} 

 In Fig.~\ref{cmr}, we show the colour magnitude relation (CMR) for
 all the spectroscopic galaxy members ($r\!\leq\!17.77$) found in this
 region in the SDSS DR7, and those that are also detected at 24\m\
 in the {\em Spitzer}/MIPS observations.  For
 comparison, the 24\m\ galaxies found in Abell~1367 are also
 shown. We only use galaxies brighter than $r\!=\!15.5$ to fit the CMR.
 The CMR is of the form $g\!-\!r\!=\!1.244\!-\!0.032\,r$ for
 all the Coma galaxies, where the mean absolute
 deviation from the relation is 
$\!\pm\!0.064$ mag.

We repeat this analysis in Fig.~\ref{ssf} for the near-infrared (NIR)
band of SDSS ($z$-band), and the 24\m\ MIPS band. The galaxies
classified as red and blue on the optical colour-magnitude diagram,
split into two separate classes around $(24\!-\!z)= -6$ mag on the
plot of the near/mid IR colour and magnitude as well. Although we
overplot the lines of constant SFR in Fig.~\ref{ssf}, according to the
empirical relation given by \citet{calzetti}, it is important for the
reader to consider that the 24\m\ flux alone is an accurate SFR
indicator only for the late-type, dusty star-forming galaxies (we
return to this issue below).  The SDSS $z$-band, centred
at $\sim$\,9000\AA, is a good measure of the light from evolved
stars, and hence the stellar mass of a galaxy.  The top axis shows the
stellar mass of galaxies estimated using the relation $\log
M^*\!=\!-0.306\!+\!1.097(g-r) \!-\!0.1\!-\!0.4(M_r\!-\!5\log\,h
\!-\!4.64)$, from  \citet{bell03}.  On the other hand, the 24\m\ MIPS band in
the MIR is a good proxy for the dust-processed light. This makes the
$(24\!-\!z)$ colour an excellent approximation for the value of the 
specific star formation rate (SSFR or SFR/M$^*$) of galaxies.
 
 Figs.~\ref{cmr} and \ref{ssf} together show that the 24\m\ detected
 red sequence galaxies in Coma have consistent optical and MIR
 colours.
 We note that the optical-MIR colours of the red sequence galaxies are
 not consistent with those expected from photospheric emission from
 old stellar populations, with an excess emission always apparent at
 24\m. The {\em Spitzer} Infrared Spectrograph observations of
 early-type galaxies in Virgo and Coma clusters show that the diffuse,
 excess emission, apparent at 10--30\m\ in these galaxies, is due to
 silicate emission from the dusty circumstellar envelopes of
 mass-losing evolved AGB stars
 \citep{bressan,clemens}. The strength of this silicate emission is a
 slowly declining function of stellar age \citep{piovan}, and persists
 even for very evolved stellar populations ($>\!10$\,Gyr). The
 optical-MIR colours of Virgo and Coma galaxies have been found to be
 consistent with such old stellar populations \citep{clemens}.

 \begin{figure} 
 \centering{{\rotatebox{270}{\epsfig{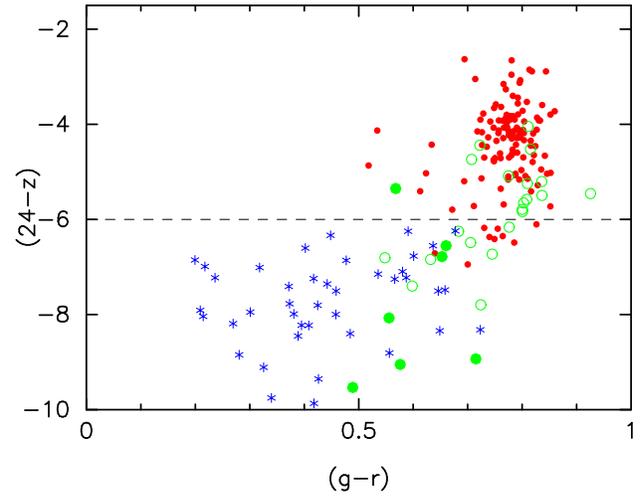}}}}
 \caption{This colour-colour diagram shows that the galaxies identified
   as AGN ({\it{green circles}}) and star-forming ({\it{blue stars}}),
   on the basis of the BPT diagram according to their optical spectra
   (see Fig.~\ref{bpt}), occupy different regions on the near/mid IR
   magnitude-colour diagram. The {\it{solid}} and {\it{open green
       circles}} represent AGN classified on the BPT diagram
   and by the \citet{miller03} criterion respectively (see text). The
   passive galaxies ({\it{red points}}) detected at 24\m\ are
   concentrated in a small region on the top right of the diagram. The
   horizontal line at $(24\!-\!z)= -6$~mag marks the boundary between these
   two classes in  the $z$-$(24\!-\!z)$ IR colour magnitude
   space (Fig.~\ref{ssf}). }
 \label{col}
 \end{figure}

 The blue galaxies in Fig.~\ref{cmr} have a wide spread in both colour
 and magnitude.  The inhomogeneity of this class of galaxies becomes
 even more clear in Fig.~\ref{col}, where we plot the optical
 $(g\!-\!r)$ colour against the near/mid-IR $(24\!-\!z)$ colour. Just
 as in Fig.~\ref{cmr}, the passive red galaxies cluster in a small
 region of the colour-colour space, but the blue galaxies and
 (optical) AGNs span a wide range along both the axes. We note the
 clear separation of star-forming (blue stars) and passive (red
 points) galaxies in $(24-z)$ colour. The horizontal dashed line in
 Fig.~\ref{col} indicates our empirically chosen criterion to separate
 the two populations about $(24-z)=-6$.  Interestingly, although the
 AGN and star-forming galaxies (green and blue symbols respectively)
 are classified on the basis of their optical spectra on the BPT
 diagram \citep[][also see \S\ref{sf}]{bpt81}, they occupy distinct
 regions in this plot of optical vs optical-IR colour, as well.

 \begin{figure}
 \centering{
 {\rotatebox{0}{\epsfig{file=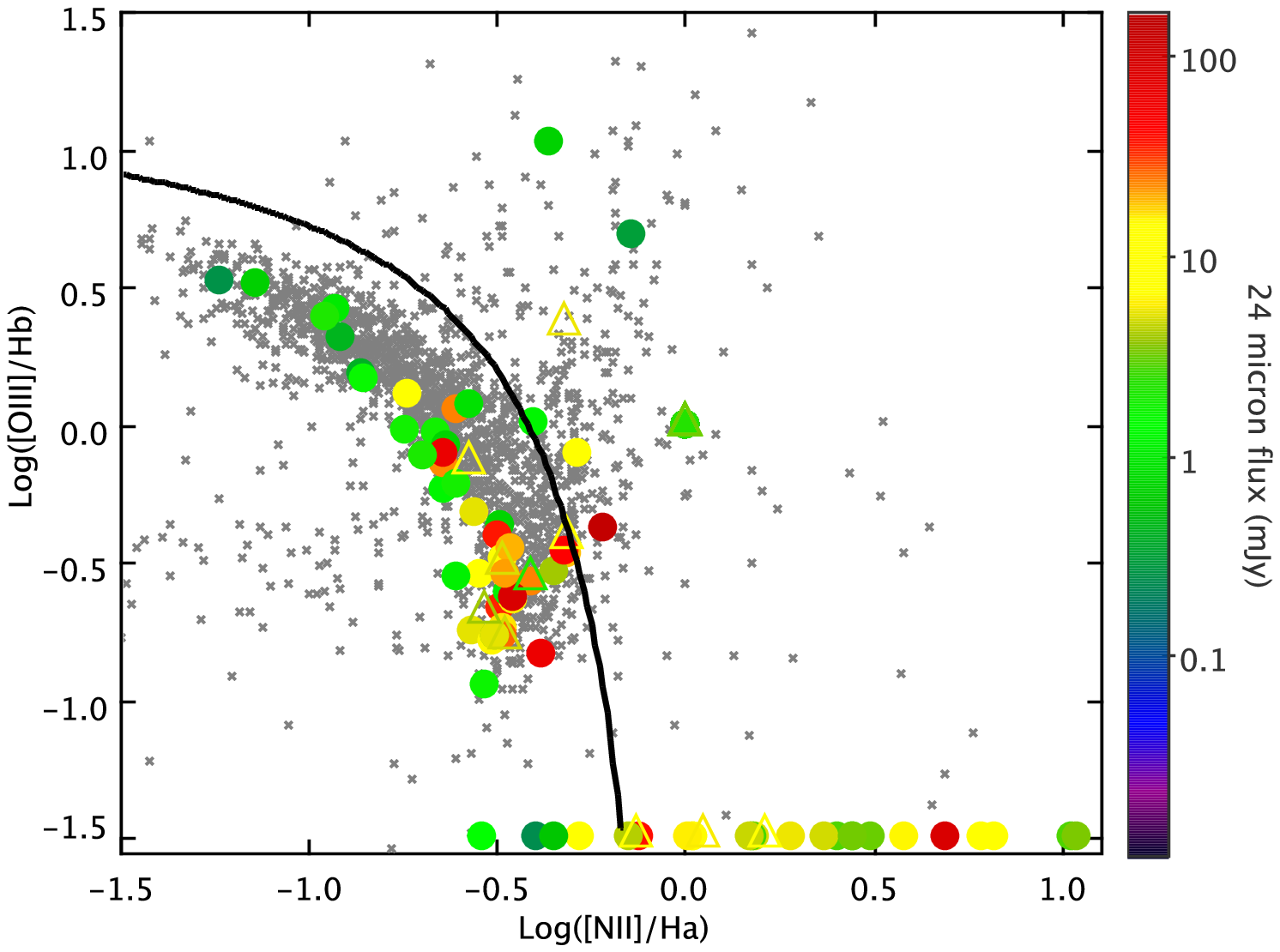,width=9.5cm}}}}
\caption{The BPT diagram \citep{bpt81} for the 24$\mu$m detected
  galaxies in the Coma cluster ({\it{filled circles}}) and in
  Abell~1367 ({\it{open triangles}}), colour coded by their 24\m\
  flux. We also plot all the galaxies within $\pm$ 2,000 km~s$^{-1}$
  of Coma or Abell~1367, in the 500 square degree  
supercluster region, but not detected at 24\m, as {\it
    grey crosses }.  The points plotted in a line at
  log[OIII]5007/H$_{\beta}\!=\!-1.5$ are the galaxies which have no
  detected [OIII] and/or H$_{\beta}$ emission. It is interesting to
  observe that a large fraction of the Coma galaxies without [OIII]
  and/or H$_\beta$ have their [NII]/H$_\alpha$ flux ratios as expected
  for AGN, suggesting that these galaxies may have their nuclear
  emission obscured. We classify such galaxies as AGN if they have
  Log([NII]/H$_{\alpha}$)$>$-0.2 \citep{miller03}. }
 \label{bpt}
 \end{figure}

\subsection{Optical and 24\m\ star formation indicators}
 \label{sf}

 The understanding of the formation of stars in galaxies requires,
 among other things, the measurement of the rate at which the
 interstellar gas is converted into stars. With the development of
 appropriate technology, radio, IR and UV photometry and spectroscopy
 are increasingly being employed, in addition to the traditional
 optical observations, for the purpose of measuring the rate of star
 formation. The dust clouds surrounding the young stellar
 concentrations absorb starlight and re-radiate it at IR
 wavelengths. Thus, the SFR measured by IR measurements is accurate
 only in the optically thick limit. The observed UV radiation escaped
 from the molecular clouds which block the UV light at earlier ages,
 comes mostly from stars with ages 10$^7$\,-\,10$^8$ yrs
 \citep{calzetti05}.  H$\alpha$ emission is produced only in the first
 few million years from the most massive stars \citep{leitherer}.
 Indeed, a combination of UV, H$_\alpha$ and IR observations is
 required to give a full measure of the obscured and unobscured star
 formation.  In this section, we attempt to combine the direct
 (optical) and the obscured (IR) radiation from the Coma and
 Abell~1367 galaxies to understand the process of star formation
 across this supercluster.
 
 The 24\m\ IR emission in galaxies can result from dust heated by
 young massive star clusters as well as the AGN. Since our MIPS data
 comes from the densest regions at the core of the Coma cluster
 (Fig.~\ref{scl}), which is a favourable environment for AGN hosts, it
 is important to investigate the origin of the 24\m\ emission in these
 galaxies. To do so, in Fig.~\ref{bpt} we plot the usual ratios of
 optical emission linewidths ([OIII], H$_\beta$, [NII] and
 H$_\alpha$), known to distinguish star forming galaxies from those
 dominated by AGN, in what is popularly known as the BPT diagram
 \citep{bpt81}.
   
 In Fig.~\ref{bpt}, we show all the emission-line galaxies found in the
 $\pm$\,2,000 km~s$^{-1}$ redshift slice around Coma and/or Abell 1367,
 in the $\sim$500 square degrees Coma supercluster region (see
 Fig.~\ref{scl}), along with the galaxies detected at 24\m\ (MIPS)
 in Coma (circles) and Abell~1367 (open triangles).  Although we find
 a significant number of 24\m\ detected galaxies with [NII]/H$_\alpha$
 ratios suggestive of AGN, this does not mean that the 24\m\ emission
 is predominantly due to AGN.
 \citet{goulding09} have tried to estimate the contribution of the AGN
 component to the IR emission from galaxies using Spitzer/IRS
 spectroscopy.  Although only a few galaxies in their sample have IR
 fluxes produced predominantly by AGN, they also find that a
 substantial fraction of AGN are optically obscured, consistent with
 the results presented here.
 
 We find that most of the galaxies which have measured values of all
 four emission lines are dusty star forming galaxies. However,
 interestingly, a large number of galaxies detected at 24\m\ in Coma
 that do not show emission in [OIII] and/or H$_\beta$, do have a
 [NII]/H$_\alpha$ flux ratio consistent with the presence of an
 AGN. Such galaxies can be classified as AGN if
 $\log$ ([NII]/H$_{\alpha}$)$>$-0.2 \citep{miller03}.
    
 \begin{figure}
 \centering{
 {\rotatebox{0}{\epsfig{file=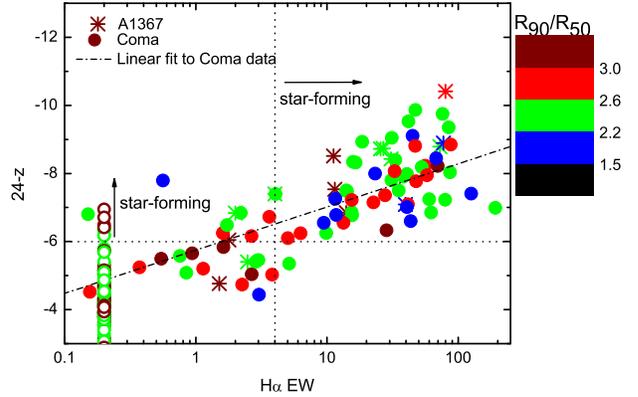,width=9cm}}}}
\caption{The ($24\!-\!z$) colour as a function of H$_{\alpha}$ EW for
  Coma ({\it{filled circles}}) and Abell 1367 ({\it{stars}}) galaxies
  respectively. The straight line represents the linear fit to
  galaxies in the Coma cluster. Though there is a considerable
  scatter, the correlation between the two quantities is
  significant. The dotted lines represent the lower limits adopted to
  select star-forming galaxies using (H$_\alpha$ EW=4\AA) and ($24\!-\!z$)
  colour\,$=\!-6.0$ (see Fig.~\ref{ssf}). For completeness, we also show the 24\m\
  detected galaxies that do not have H$_\alpha$ emission in SDSS spectra at
  EW(H$_\alpha$)\,$=\!0.2$\AA~({\it{open circles}}). The 24\m\ emission in these
  galaxies is likely to come from the evolved AGB stars. }
 \label{ha-24z}
 \end{figure}
 
 The EW of the Balmer lines, especially that of the H$_\alpha$
 emission line, has been extensively used to estimate the current
 optical SFR of galaxies. It has been shown in the literature that the
 24\m\ flux is a good measure of dust-processed star light
 \citep[\eg][]{calzetti}. But, the 24\m\ observations used in this
 work cover the very central regions of the Coma cluster, and
 Abell~1367, which comprises mostly of early-type galaxies.  In order
 to test the correspondence between the direct and dust-processed SFR
 tracers, in Fig.~\ref{ha-24z} we plot the $(24\!-\!z)$ colour as a
 function of spectroscopically measured EW(H$_\alpha$). We also
 colour-code the symbols according to the concentration parameter of
 the galaxies (ratio of the Petrosian radii R$_{90r}$/R$_{50r}$ from
 the SDSS photometric catalogue, which is an indicator of morphology).
 Passive spirals often have concentration indices consistent with early-types
 \citep[R$_{90r}$/R$_{50r}\!>\!2.6$; also see][]{mahajan09a}. But
 Fig.~\ref{ha-24z} shows that both the spiral and spheroidal galaxies
 are uniformly distributed along both the axes, implying that the emission-line
 galaxies detected at 24\m\ at the core of Coma and Abell 1367 are not
 dominated by galaxies of any particular morphological type (as
 quantified by the concentration index).

 Due to the dominance of early-type galaxies in our 24\m\ sample, it is
 not surprising that only a small fraction of galaxies classified as
 star-forming on the basis of their $(24\!-\!z)$ colour
 (Fig.~\ref{ssf}), show no signs of current star formation in their
 optical spectra. 
 For completeness, at EW(H$_\alpha$)\,$=\!0.2$\AA~we plot all the 24\m\
 detected galaxies that do not show optical emission in H$_\alpha$. As
 expected, most of these galaxies have $(24\!-\!z)$ colours and concentration
 parameters consistent with those of quiescent early-type galaxies. This
 suggests that their 24\m\ emission is primarily due to the contribution
 from AGB stars, which (unlike young stars) do not produce H$_\alpha$ emission. 
 The good overall correspondence between the
 photometric $(24\!-\!z)$ colour and spectroscopic EW(H$_\alpha$)
 makes the $(24\!-\!z)$ colour a good candidate for comparing the star
 formation activity in nearby galaxies, in the absence of optical spectra.

 \subsection{Varying fractions of star-forming galaxies with SFR
   tracer: implications for the Butcher-Oemler effect}
 \label{fractions}

 \begin{figure*}
 \centering{
 {\rotatebox{270}{\epsfig{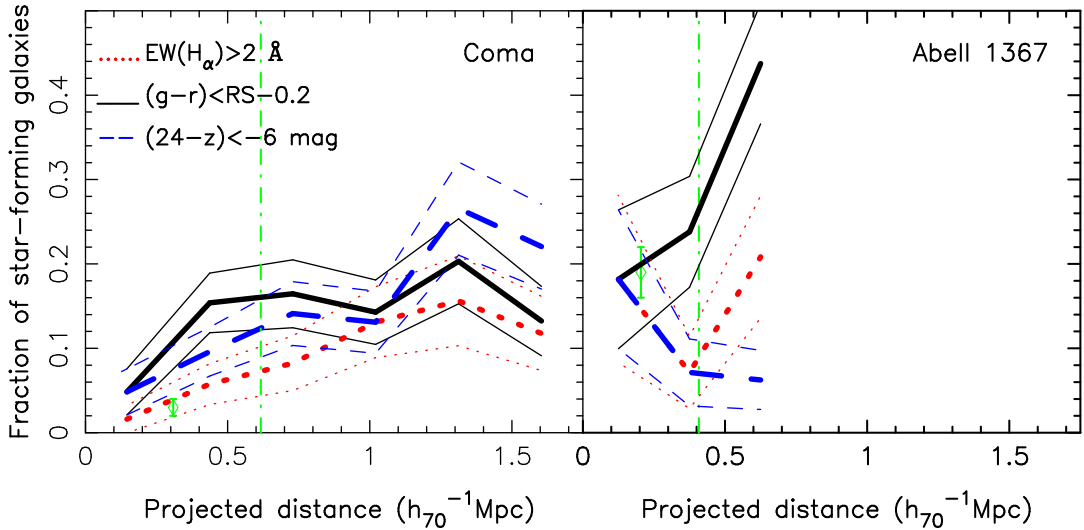}}}}
\caption{The fraction of star-forming galaxies as a function of
  (projected) cluster-centric radius for Coma in the {\bf{Left}} panel
  and Abell 1367 on the {\bf{Right}}. In both panels, the {\it{solid
      black line}} shows the radial variation in the fraction of
  star-forming galaxies, selected using the $(g\!-\!r)$ colour
  threshold, the {\it{dotted red line}} makes use of the H$_\alpha$ EW
  ($\geq$2\AA) and the {\it{dashed blue line}} represents the same,
  for star-forming galaxies chosen by the $(24\!-\!z)\!<\!-6$
  criterion. The corresponding {\it{thin curves}} show the $\pm\!1\sigma$
 Poisson scatter. In all cases, all the spectroscopically
  identified galaxy members with $M_{z}\!<\!-20.82$
  ($M^*_{z}\!+\!1.5$; see text), and l.o.s.  velocity $\pm$ 3,000
  km~s$^{-1}$ of Coma and/or Abell 1367, are used to estimate the
  fractions.  For Abell~1367, we show galaxies only within the region,
  of scale $\sim$\,0.75 $h^{-1}_{70}$~Mpc, for which the 24\m\ data is
  available.  This diagram shows the importance of taking account of
  the star formation activity  used to quantify and understand
  evolutionary trends, such as the Butcher-Oemler effect. For
  comparison, the R$_{30}$ radius adopted by \citet{boe}, and the
  fraction of blue galaxies found in Coma and Abell 1367, are also
  shown by the vertical {\it{dot-dashed lines}} and {\it{open green
      diamonds}} respectively.}
 \label{sf-frac}
 \end{figure*}

 Photometric colours and EWs of emission lines like [OII] and
 H$_\alpha$ have been extensively used for studying the evolution of
 galaxies in time and across the sky. In one such pioneering work,
 \citet[][BO84]{boe} found that clusters at moderate to high redshifts contain
 an `excess' of blue galaxies, relative to their local counterparts.
 They estimated the blue fraction by considering galaxies found within
 R$_{30}$ (radius containing 30\% of all red sequence galaxies), for
 which the optical broadband colours are bluer by at least 0.2 mag
 than that of the red sequence galaxies.  For the redshift regime of
 Coma ($\leq\!0.1$), BO84 found a uniform blue fraction within
 R$_{30}$. Since then, several such studies
 have sought to quantify 
and validate the Butcher-Oemler effect for different samples
 of clusters.  Some of these 
define their samples in a way similar to that of BO84,
 and obtain similar results
 \citep[\eg][]{margoniner00}.  

Using panoramic MIR data for 30
 clusters (including both Coma and Abell 1367) over $0\!<\!\rm z\!<\!0.4$,
 \citet{chris09} are able to reproduce the Butcher-Oemler effect using
 a fixed limit in $L_{IR}$ $(5{\times}10^{10}L_{\odot})$ (equivalent
 to a fixed SFR of 8\,M$_{\odot}$yr$^{-1}$), but show that the
 Butcher-Oemler effect can be largely explained as a consequence of
 the {\em cosmic} decline in star formation \citep{lefloch,zheng07}.
 In this case, the blue galaxies in clusters are those that are recently
 accreted from the field \citep[accretion has occurred at a relatively constant
 rate since $\rm z\!\sim\!0.5$;][]{berrier09}, but since the global
 level in star formation among these galaxies has declined, a smaller
 fraction of the infalling population is classed as blue (which
 assumes a non-evolving level of star formation), resulting in the
 observed Butcher-Oemler effect.

 Elsewhere \citep{bnm00,ellingson01,depropris04}, studies going out to
 several multiples of the cluster-centric radius, scaled by R$_{200}$,
 show the effect of the chosen aperture size on the blue fraction.  It
 has been shown that the observed gradual radial trend of $f_{SF}$ is
 consistent with a simple infall scenario, whereby the star-forming
 galaxies are infalling field galaxies, which are then quenched
 rapidly upon their first passage through the cluster core
 \citep[\eg][]{balogh00,ellingson01,chris09}. By comparing the
 photometric colour with the spectroscopically determined SSFR,
 \citet{mahajan09a} show that the presence of metal-rich stellar
 populations in low redshift cluster galaxies can also influence the
 evolutionary trends such as the Butcher-Oemler effect, if they are
 studied only using galaxy colours.  Several recent studies indicate
 that a non-negligible fraction of red sequence galaxies show signs of
 ongoing star formation from their optical spectra and/or broad-band
 colour, and that a robust separation of passive and star-forming
 galaxies requires mutually independent data %multi-wavelength data
 \citep[\eg][]{bildfell}.  In this work, we  show that in the
Coma Supercluster,
 the value of $f_{SF}$ not only varies with the
 cluster-centric aperture used for measuring the blue fraction, but
 can also be severely effected by the SFR tracer employed.
 
 In Fig.~\ref{sf-frac} we plot the fraction of non-AGN, star-forming
 galaxies found using 3 different criteria: (i) IR colour
 [($24\!-\!z)\!\leq\!-6$ mag; Fig.~\ref{ssf}], (ii)
 EW(H$_\alpha$)$\geq\!2${\AA}~and, (iii) the photometric colour
 $(g\!-\!r)$ is bluer than that of the fitted red sequence
 (Fig.~\ref{cmr}, by more than the mean absolute deviation, for all
 the Coma and Abell 1367 galaxies. BO84 had included galaxies brighter
 than $M_{V}\!=\!-20$ (H$_{0}\!=\!50$ km~s$^{-1}$Mpc$^{-1}$) in their
 sample.  To make a fair comparison with the work of \citet{boe},
 without adding uncertainties by using empirical relations to convert
 magnitudes from one passband to the other, in Fig.~\ref{sf-frac} we
 choose to only include galaxies brighter than $M^{*}_{z}\!+\!1.5$
 \citep[$M^{*}_{z}\!=\!-22.32$;][]{blanton01} for calculating the
 fractions. Also, note that {\it all} the spectroscopic galaxy members
 are used to calculate the fractions in each radial bin.
 
 These distributions show that, by taking into account the obscured
 star formation estimated from the 24\m\ flux, the fraction of
 star-forming galaxies can dramatically vary at any given radius from
 the centre of the cluster. In both clusters, the `blue' fraction
 ($f_b$) is higher, and flattens at lower cluster-centric radii, than the 
 $f_{SF}$ estimated using the
 EW(H$_\alpha$), evidently showing a significant contribution of the
 post-starburst galaxies to the fraction $f_b$ (also see
 Section~\ref{ka}). We note that this trend may vary if the dusty red
 galaxies have a non-negligible contribution
 in building the red sequence. But as seen in Fig.~\ref{cmr}, in the
 Coma cluster, this does not seem to be the case.

 It is interesting to note that unlike Coma, Abell~1367 has an
 increasing fraction of blue galaxies with cluster-centric radius
 (black lines), but an inverse trend emerges when the fractions are
 measured using the $(24\!-\!z)$ colour (blue lines) or the H$_\alpha$
 EW (red lines). This implies that some of the blue galaxies outside
 the core of Abell~1367 are post-starburst galaxies (see
 Fig.~\ref{coma-gals}). This result supports the results obtained in
 the more general work of \citet{mahajan09a}, who show that using a
 single galaxy property, such as the broadband colour, is not an
 appropriate way of quantifying evolutionary trends like the
 Butcher-Oemler effect.
         
 \section{Discussion}
 \label{discussion}

 In this paper we set out to understand the relationship between the
 star formation activity in galaxies, as depicted by an assortment of
 indicators encompassing the optical and 24\m\ mid-IR wavebands, and
 their immediate and global environment, in the Coma supercluster.  A
 wide range of local and global environments of galaxies, and a
 uniform optical coverage across the entire $\sim\!500$ square degrees
 of sky, make this supercluster an ideal laboratory for examining the
 environmental dependence of galaxy properties. We discuss below the
 implications of the results from our analysis presented in
 \S\ref{coma-sf} and \S\ref{results}.

%%%%%%%%%%%%%%%%%%%
 \subsection{The spatial and velocity distribution of galaxies detected
  at 24\m}
 \label{mips}

 \begin{figure}
 \centering{
 {\rotatebox{270}{\epsfig{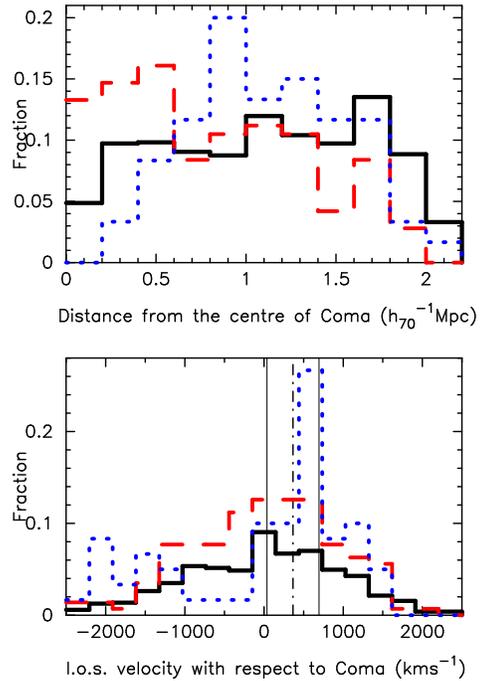}}}}
\caption{{\bf{(Top panel):}} The 
  distribution of cluster-centric distance for all the
  spectroscopically identified  
 galaxies in a $\pm$3,000 km~s$^{-1}$ slice in the Coma cluster
({\it{solid black}} histogram).
  The {\it{dashed red}} and {\it{dotted blue}} histograms show the
  same for the 24\m\ MIPS galaxies, divided into
 red and blue galaxies on the basis of the
  $(g\!-\!r)\!-\!r$ colour-magnitude plot (Figs.~\ref{cmr} \&
  \ref{ssf}). The horizontal axis is limited by the coverage of the
  MIPS field. Interestingly, even though the red 24\m\ galaxies are
  uniformly distributed within a $\sim$2~Mpc radius, the blue galaxy
  population seems to peak away from the cluster core at 1--1.5~Mpc from
  the centre. {\bf{(Bottom panel):}} These histograms represent
  the same galaxies as in the upper panel, but for the
  l.o.s. velocity of galaxies, relative to the mean redshift of the
  Coma cluster.  Intriguingly, while the red galaxies follow the
  distribution of {\it all} the spectroscopic galaxy members of the Coma
  cluster, the blue 24\m\ galaxies show a remarkable peak around the
  velocity of the galaxy group NGC\,4839, shown here as the
  {\it{dot-dashed}} line, with $\pm \sigma_v\!=\!329$\,km~s$^{-1}$
  \citep{colless96}. }
 \label{radius}
 \end{figure}

 The 24\m\ data for the Coma cluster extend out to a few times its
 core radius (Fig.~\ref{sf-frac}), allowing us to analyse the spatial
 (sky and velocity) distribution of the galaxies detected at 24\m,
 relative to all the spectroscopic members found in the SDSS. In order
 to do so, in Fig.~\ref{radius}, we plot the distribution of all the
 spectroscopic galaxies, and the (optically) red and blue galaxies
 detected at 24\m\ respectively. As discussed above (Figs.~\ref{cmr}
 and \ref{ssf}), in Coma both the 24\m\ and optical colours mostly
 segregate the same galaxies into `red' and `blue'
 ones. Fig.~\ref{radius} (top panel) shows that, of the 24\m\ detected
 galaxies, the distribution of the (optically) red ones is similar to
 that of all the galaxies, but the (optically) blue ones tend to peak
 $\gtrsim$1.0 $h^{-1}_{70}$Mpc from the centre of Coma. % However, a
% Kolmogorov-Smirnoff (K-S) test suggests that the differences in these
% distributions are statistically insignificant.
   
 In Fig.~\ref{radius} (bottom panel), we plot the distribution of the
 line of sight (l.o.s.) velocities of galaxies, with respect to the
 mean velocity of the Coma cluster (called the `relative velocity'
 hereafter), for the same three sets of galaxies. The distribution of
 the relative velocity of `all' and (optically) red galaxies are
 statistically similar. But the (optically) blue 24\m\ galaxies show a
 bimodal distribution, with a large fraction of one mode concentrated
 around relative velocity $\sim\,600$ km~s$^{-1}$. A Kolmogorov-Smirnoff
 (K-S) test suggests
 that the probability of the parent distribution of the relative
 velocities of red and blue galaxies being sampled from the same
 population is $p\!=\!8.023\times10^{-5}$.  The same comparison
 between the relative velocity distribution of `all' and blue galaxies
 gives $p\!=\!0.005$, suggesting that it is highly unlikely that the
 blue and red (or `all') galaxies are drawn from the same parent
 distribution (Fig.~\ref{radius}).  This leads us to conclude that a
 non-negligible fraction of the (optically) blue 24\m\ galaxies within
 $2\,h^{-1}_{70}$~Mpc of the centre of the Coma cluster are
 star-forming galaxies which may belong to the substructure associated
 with NGC\,4839 (see \S\ref{ngc4839} for further discussion).

 The l.o.s. velocity distribution of the (optically) blue 24\m\ galaxies 
 is highly non-Gaussian, with an excess at $\sim\!600$ km s$^{-1}$, suggesting
 that many of them have recently entered the cluster. If these blue 24\m\
 galaxies were virialized, a Gaussian distribution centered on $0$,
 like that seen for the red (red dashed histogram) and `all' (solid black histogram)
 galaxies would be expected. 

 The current episode of star formation in these galaxies can be
 attributed to the environmental impact of the cluster's ICM
 \citep[\eg][]{poggianti04}, or the enhanced galaxy density in the
 infall region \citep[\eg][]{mrp10}. This result is consistent with
 the findings of \citet{caldwell97}, who analysed the spectra of
 early-type galaxies in 5 nearby clusters, including Coma, and found
 that in 4 of the 5 clusters, the early-type galaxies show 
 signatures of recent star formation in their spectra (also see \S\ref{ka}).

 \subsection{The distribution  of k+A galaxies}
 \label{ka} 

 \begin{figure*}
 \centering{
 {\rotatebox{270}{\epsfig{file=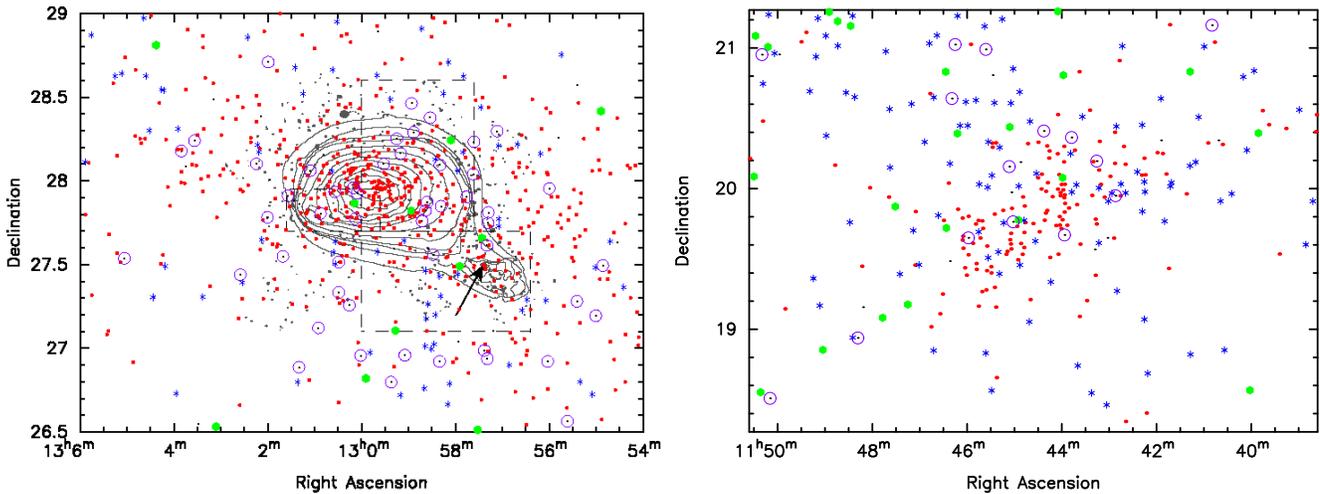,width=6.5cm}}}}
 \centering{
 {\rotatebox{270}{\epsfig{file=figure-12b,width=6.5cm}}}}
\caption{{\bf{(Left panel):}} The distribution of the passive {\it{(red
      dots)}}, AGN host {\it{(green points)}}, star-forming {\it{(blue
      stars)}} and the post-starburst (k+A) {\it{(purple circles)}}
  galaxies in $\sim\!5.0\!\times\!4.2$\,$h^{-1}_{70}$Mpc region
  surrounding the centre of Coma. The contours show X-ray emission
  from a {\it{XMM-Newton}} EPIC/PN observation. 
 As can be noticed, not only the
  star-forming, but also the k+A galaxies seem to avoid the dense
  cluster centre. Also, the presence of k+A galaxies out to almost
  twice the virial radius from the centre shows how strong the impact
  of the `global' cluster environment is on the evolution of galaxies
  in the vicinity of massive structures.
  {\bf{(Right panel):}} Same as above, but for the
  $5.0\!\times\!5.0$\,$h^{-1}_{70}$Mpc region surrounding the centre
  of Abell 1367. It is interesting to see that in contrast with Coma,
  most of the k+A galaxies in Abell 1367 seem to be aligned along the
  direction of the filament feeding it from the direction of Coma. We
  also note that the (optical) AGN in both the clusters are mostly
  found in the direction of the filament connecting Coma and Abell
  1367. }
 \label{coma-gals}
 \end{figure*} 
 
 \begin{figure}
 \centering{
 {\rotatebox{270}{\epsfig{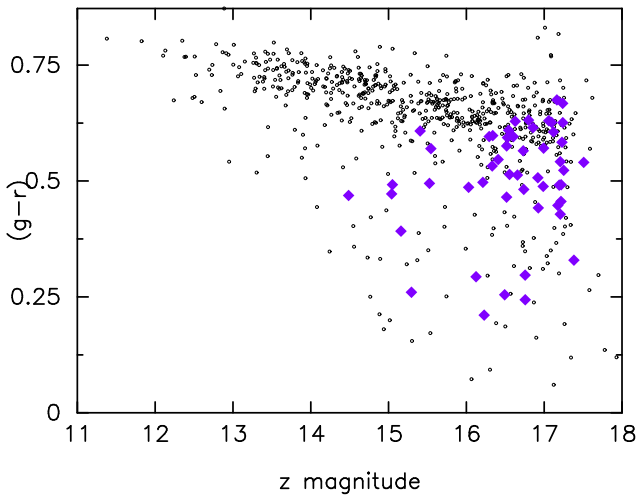}}}}
\caption{ The $(g\!-\!r)\!-\!z$ colour-magnitude diagram of all the
  spectroscopically identified galaxies {\it{(black points)}} in the
  $\sim\!5.0\!\times\!4.2$\,$h^{-1}_{70}$Mpc region surrounding the
  centre of the Coma cluster (shown in Fig.~\ref{coma-gals}).
  The k+A galaxies {\it{(purple diamonds)}} are mostly blue
  dwarfs. These observations confirm that the low-mass galaxies are
  the first ones to experience and reflect the impact of rapid change in
  their local environment.  }
 \label{ka-cmr}
 \end{figure}
  
 The post-starburst, or k+A galaxies, as they are popularly known in
 the literature, are passive galaxies that have recently (1-1.5 Gyr)
 experienced a strong burst of star formation. These galaxies are very
 crucial in understanding the impact of environment on various galaxy
 properties, especially their SFR. The spectrum of a k+A galaxy shows
 strong absorption in H$_\delta$, but little or no emission in
 H$_\alpha$.
 
 In this work we make use of the SDSS DR7 spectroscopic galaxy
 catalogue ($r\!\leq\!17.77$) for identifying the k+A galaxies
 (EW(H$_{\delta})\!<\!-3$\AA~\& EW(H$_{\alpha})\!<\!2$\AA) in the Coma
 supercluster (Please note that throughout this work, negative values
 of EW indicate absorption). In Fig.~\ref{coma-gals} we show these k+A
 galaxies together with the passive, star-forming and AGN galaxies in
 and around the Coma and Abell 1367 cluster respectively. As can be
 easily seen, the k+A galaxies preferentially avoid the dense region
 in the cluster core but inhabit the surrounding infall regions out to
 $\sim\!5$\,$h^{-1}_{70}$Mpc (almost twice the virial radius).  In
 Fig.~\ref{coma-gals} we have also over-plotted the contours %for the
 of intensity from the mosaicked 0.5--2~keV image of the core of Coma
 cluster from XMM-Newton EPIC/PN observations \citep{fino03}, kindly
 supplied to us by A.~Finoguenov.  We note that there are almost no
 star-forming galaxies in the X-ray emitting region of the core.  This
 observation further strengthens the argument that the changes in
 environment, experienced by a galaxy on the outskirts of clusters,
 play a key role in modulating the properties of galaxies, especially
 the dwarfs \citep[\eg][]{porter07,porter08}.
 
 By analysing deep ($M_{B}\!\lesssim\!-14$) photometric and
 spectroscopic optical data for 3 regions in Coma (two near the centre
 and one near NGC\,4839, each $\sim\!1\!\times\!1.5$ Mpc in size),
 \citet{poggianti04} found that $\sim$\,10\% of the cluster's dwarf
 ($M_{V}\!>\!-18.5$) galaxies have post-starburst spectra. They also
 used the results from the X-ray analysis of \citet{neumann03} to show
 that the k+A galaxies in the Coma cluster are likely to be a result
 of the interaction between the Coma cluster and the adjoining
 NGC\,4839 galaxy group (see \S\ref{ngc4839}). In
 Fig.~\ref{coma-gals}, we overplot as dashed rectangles, the
 approximate location of the substructures found by
 \citet{neumann03}. As has been shown by \citet[][their
 fig.~6]{poggianti04}, on the western side of the Coma cluster, the
 k+A galaxies seem to lie along the X-ray substructure \citep[also see
 fig.~2 of][]{neumann03}. We demonstrate this by overplotting contours
 of X-ray intensity in Fig.~\ref{coma-gals}.

 However, it is also interesting to note that the southern substructure,
 apparent from the X-ray contours, is almost devoid of k+A galaxies,
 but has a stream of star-forming dwarf galaxies flowing towards the
 cluster core (Fig.~\ref{mean_ha}; also see \S\ref{coma-scl}). Contrary to
 \citet{poggianti04}, by taking into account the extended region surrounding
 the Coma cluster, we find that the k+A galaxies preferentially avoid the
 cluster core but their spatial distribution does not show any correlation
 with the substructure evident in the X-ray emission (Fig.~\ref{coma-gals}).

 In Fig.~\ref{ka-cmr} we examine the position of the k+A galaxies in
 the Coma cluster in the plot of $(g\!-\!r)$ colour vs $z$ magnitude.
 Fig.~\ref{ka-cmr} show that almost all the k+A galaxies in Coma are
 dwarfs ($z\!\lesssim\!15$), suggesting that either only the infalling
 dwarf galaxies are rapidly quenched, or the SFR-$M^*$ relation
 \citep[\eg][]{feulner} dilutes the post-starburst signature in
 massive galaxies. 
 Since dwarf k+A galaxies in the Coma supercluster are found on the
 outskirts of Coma and Abell~1367 clusters, and occasionally in galaxy
 groups embedded elsewhere in the large-scale structure (LSS), this
 might suggest that dwarf galaxies falling into deeper potentials are
 more likely to show k+A features, while those being assimilated into
 galaxy groups may be quenched on a longer time-scale. We will probe
 this in greater detail in a later paper.

 The handful of galaxies with very blue colours ($(g\!-\!r)\!<\!0.5$;
 Fig.~\ref{ka-cmr}) are the k+A dwarfs in which the episode of
 starburst could have ended $\sim\!300$\,Myr ago, unlike the more
 evolved red post-starburst dwarfs \citep{poggianti04}.  By studying
 spectra of early-type galaxies ($M_{b}\!<\!16.7$) in Coma,
 \citet{caldwell97} also found that $\sim$\,15\% of these galaxies
 show signatures of recent or ongoing star formation. Based on this
 result, \citet{caldwell97} suggested that the present day clusters
 act as a catalyst for galaxy evolution, though at a reduced level as
 compared to their high redshift counterparts.
 
 \begin{table}
 \begin{minipage}{\linewidth}
 \centering{
 \caption{Dwarf ($z\!>\!15$ mag) galaxies in the Coma supercluster}
 \label{tbl:ka-nos}
 \begin{tabular}{|c|c|c|c|}
 \hline
      &  Total  & Star-forming & k+A \\
 \hline

 Coma         &   438   &   55         &  50  \\
 ($\sim\!5\!\times\!4.2$\,Mpc$^2$; Fig.~\ref{coma-gals}) & & & \\ %\hline
 Abell 1367   &   391   &  203         &  19  \\
 ($\sim\!5\!\times\!5$\,Mpc$^2$; Fig.~\ref{coma-gals}) & & & \\ %\hline
 Supercluster           &  1106   & 667  & 23  \\
 (excluding regions mentioned above) & & & \\
 \hline
 \end{tabular}}
 \end{minipage}
 \end{table}

 Given the vulnerable nature of dwarf galaxies, it is not surprising
 that a large fraction of them are the first ones to reflect the
 impact of rapid changes in their environment during the transition
 from filament to cluster. This contributes to dwarf galaxies having a
 better-defined SF-density relation, relative to their more massive
 counterparts (Fig.~\ref{mean_ha}). 
 Table~\ref{tbl:ka-nos} shows that
 11.4\% of dwarfs in Coma, 4.8\% in Abell 1367 and 2.1\% in the
 neighbouring supercluster region have spectra with k+A features. This
 evidently shows that the mechanisms responsible for quenching star
 formation in dwarf galaxies and rapidly transforming them to passive
 galaxies via the post-starburst phase, are strongly dependent on the
 cluster potential.  This result is also in agreement with the
 observation of excessive red, dwarf ellipticals (dEs) in the Coma
 cluster \citep{jenkins07}. In Table~\ref{tbl:ka-catalogue} we provide a
 list of all the 110 k+A dwarf ($z\!>\!15$) galaxies found in the
 entire Coma supercluster.

 \begin{table}
 \begin{minipage}{\linewidth}
 \centering{
 \caption{Catalogue of dwarf ($z\!>\!15$), 
   k+A (EW(H$_\alpha$)\,$<\!2$\,\AA~ \&
   EW(H$_\delta$)\,$<\!-3$\,\AA) galaxies in the Coma supercluster (this table is
   available in full online).}
 \label{tbl:ka-catalogue}
 \begin{tabular}{|c|c|c|c|c|c|}
 \hline
 RA       &  Dec     & Redshift & $z$  & EW(H$_\alpha$) & EW(H$_\delta$) \\
 (J2000)  & (J2000)  &        & mag    & \AA            & \AA            \\
 \hline
 195.4972 & 28.7095 & 0.0203 & 16.8559 & -1.9535 & -20.5023 \\
 194.7338 & 28.4636 & 0.0198 & 15.1591 & -2.6499 & -5.6373  \\
 174.6209 & 28.5871 & 0.0236 & 15.3615 &  1.8792 & -3.9275  \\
 192.4220 & 28.8448 & 0.0218 & 17.1838 & -1.3320 & -30.0913 \\
 192.8569 & 28.7203 & 0.0241 & 17.0417 & -1.1904 & -3.1975  \\
 193.9028 & 26.5644 & 0.0211 & 16.9923 & -1.8170 & -6.2012  \\
 194.5201 & 26.3706 & 0.0232 & 15.7774 & -1.8387 & -6.3831  \\
 195.0435 & 26.4612 & 0.0221 & 16.6272 & -2.0973 & -3.2747  \\
 189.0882 & 27.0333 & 0.0251 & 16.6488 & -1.9006 & -3.0789  \\
 194.0083 & 26.9208 & 0.0191 & 16.5278 & -1.4534 & -4.9134  \\
 \hline
 \end{tabular}}
 \end{minipage}
 \end{table} 

 \subsection{Coma and NGC\,4839}
 \label{ngc4839}

 \begin{figure}
 \centering{
 {\rotatebox{270}{\epsfig{file=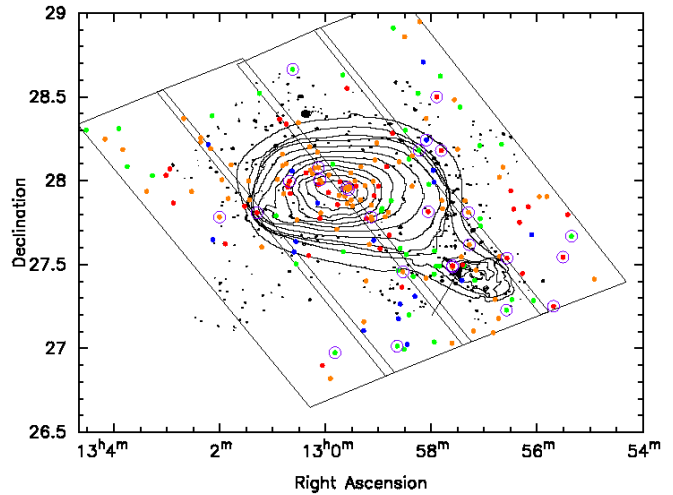,width=6.5cm}}}}
\caption{The spatial distribution of galaxies detected at 24\m\ in the
  Coma Supercluster region. The galaxies are coded by their
  ($24\!-\!z$) colour as following: red:$>\!-4$, orange:
  $-4\leq\!(24\!-\!z)\!<\!-6$, green:$-6\leq\!(24\!-\!z)\!<\!-8$ and
  blue: $\leq\!-8$.  The {\it{purple circles}} are the blue 24\m\ galaxies
  moving at velocities similar to that of the NGC\,4839 galaxy group
  \citep[\ie~having l.o.s.
  velocity\,$=\!7339\!\pm\!329$\,km~s$^{-1}$;][]{colless96}. We note
  that 19 of these 26 galaxies have ($24\!-\!z$) colour corresponding
  to that of the passive galaxies (\ie~$(24\!-\!z)\!\leq\!-6$;
  Fig.~\ref{ssf}). The contours are from a 0.5--2~keV XMM-Newton
  EPIC/PN X-ray mosaic image of the extended Coma cluster. }
 \label{coma-mips}
 \end{figure}

 As noted above, the Coma cluster is known to have significant
 substructure in the optical and X-ray maps. The most prominent X-ray
 emitting substructure is associated with a galaxy group, of which
 NGC\,4839 is the most prominent galaxy.  By combining N-body
 hydrodynamical simulations with {\it{ROSAT}} X-ray and VLA radio
 observations, \citet{burns94} remarked that the NGC~4839 group has
 already passed through the core of Coma $\sim$\,2 Gyr ago, and is now
 on its second infall \citep[also see][]{caldwell97}. However, several
 other studies based on optical spectroscopy and imaging, and X-ray
 observations conclude otherwise
 \citep{colless96,neumann03,adami05}. No consensus seems to have 
 been reached on the dynamical state of NGC\,4839. Studies supporting
 the first infall argument suggest that the NGC\,4839 galaxy group has
 a velocity dispersion ($\sigma_{cz}$) of 329 km~s$^{-1}$
 \citep[\eg][]{colless96}, while others claim it to be as high as 963
 km~s$^{-1}$ \citep[\eg][]{caldwell93}, the latter being consistent
 with the scenario where the galaxies belonging to the group have
 dispersed during their first passage through the cluster.

 Located $\sim\,1.1$~Mpc from the centre of the Coma
 cluster, the NGC\,4839 galaxy group was first detected as an
 asymmetric extension to the otherwise relaxed X-ray morphology of
 Coma \citep{briel92}.  Several authors have discovered post-starburst
 (k+A) galaxies associated with this group
 \citep{caldwell93,caldwell97,poggianti04}.  In this work, we find
 that most of the known post-starburst galaxies in Coma, such as those
 found by \citet{poggianti04}, are either not detected at 24\m, or show
 little MIR emission. We hence confirm that almost none of these
 post-starburst galaxies have a significant amount of obscured star
 formation going on in them.
     
 \citet{struck06} suggests that when a galaxy group passes through the
 core of a cluster, as suggested by \citet{burns94} for Coma and
 NGC\,4839, the group galaxies are gravitationally shocked.  In this
 scenario, enhancement in star formation activity is a natural
 consequence of an increase in galaxy-galaxy interactions among group
 galaxies.  Such a scenario has been suggested for Coma and NGC\,4839
 by \citet{neumann03}.  Fig.~\ref{coma-mips} may provide
 circumstantial evidence to support this argument.  

 In Fig.~\ref{coma-mips} we show the spatial distribution of the 24\m\
 detected Coma galaxies colour-coded by their ($24\!-\!z$) colour. The
 blue 24\m\ galaxies constituting the peak, which also corresponds to
 the mean velocity of the NGC\,4839 group (Fig.~\ref{radius}), are
 explicitly shown. 
% We note that 19 of the 26
% galaxies belonging to this peak have colours consistent with that of
% passive 24\m\ detected galaxies ($(24\!-\!z)\leq\!-6$; Fig.~\ref{ssf}).
% The probability of picking 19 red galaxies out of the 197 24\m\ detected
% galaxies (of which 143 are red) is $\sim\!8.77\!\times\!10^{-4}$,
% suggesting that the occurrence of these galaxies in a narrow velocity
% range (Fig.~\ref{radius}) may not be an event of random chance. 
% The fact that these galaxies are not concentrated close to NGC\,4839, and
% a large fraction of them are passively evolving, may indicate that
% the NGC\,4839 galaxy group has already passed through the centre of
% Coma and is now on its second infall towards the Coma centre
% \citep[also see][]{burns94,caldwell97}. However, our analysis does
% not reject a scenario where these galaxies could have been
% pre-processed within the NGC\,4839 group, before the latter's first
% infall onto Coma.
 However, poor statistics in our data do not allow us to favour any of
 the 2 scenarios associated with the NGC\,4839 galaxy group and the Coma
 cluster, namely, (i) whether the group is on it's first infall, or (ii)
 it has already passed through the core and is on it's second passage.
  
 \subsection{Coma, Abell 1367 and the filament}
 \label{coma-scl}
 
 In the hierarchical model, galaxy clusters grow by accretion and/or
 mergers with other clusters and groups. In simulations,
 galaxies are seen to be accreted along 
the network of filaments of galaxies, feeding these clusters,
along preferred directions
 \citep[\eg][]{bkp96}.  Although several
 observational techniques are now being implemented to detect and
 quantify such large-scale structures (LSS), the low surface
 density of matter, enormous spatial scale, and  projection
 effects make it difficult to observe the extent of
 these filaments of galaxies \citep{colberg07}.

 One way of exploring the evolution of the LSS is to probe its impact
 on the galaxies traversing them.  The spectacular filament crossing
 the Coma and Abell~1367 clusters is an exclusive object in the low
 redshift Universe, because it is not only traced by the spatial
 distribution of galaxies (Fig.~\ref{scl}), but has also been detected
 at radio wavelengths \citep{kim89}. One of the aims of this paper is
 to understand and interpret the difference in the evolutionary paths
 adopted by galaxies in the vicinity of the clusters, the filament and
 the groups embedded in the filament. In agreement with the scenario
 emerging in the literature \citep[][among
 others]{gray04,tanaka,smith06,chris07}, we find that the SFR-density
 relation across the Coma supercluster is much weaker for the massive
 galaxies ($z\!<\!14.5$; Fig.~\ref{sf_frac}), relative to the dwarfs
 ($z\!>\!15$; Figs.~\ref{sf_frac} \&~\ref{mean_ha}).

 Empirical studies of galaxies infalling into clusters along filaments
 suggest that an enhanced galaxy density at the cluster periphery may
 lead to a burst of star formation in them, consuming a large fraction
 of cold gas \citep[\eg][]{porter08,mrp10}.  This effect would be more
 efficient for the dwarf galaxies. In a study based on the SDSS data,
 \citet{chris07} find that the dwarf galaxies ($-19\!<\!M_r\!<\!-18$)
 residing in high density regions show a systematic reduction of
 $\sim\!30\%$ in their H$_\alpha$ emission relative to the mean of the
 sample, leading the authors to favour slow quenching of star
 formation in these galaxies \citep[also see][]{balogh04a,tanaka}.  On
 the contrary, in Coma we find that the infalling star-forming dwarf
 galaxies undergo a burst of star formation, followed by rapid
 quenching (Fig.~\ref{mean_ha}, also see Section~\ref{ka}). 

 In contrast, star formation in dwarf galaxies infalling into galaxy
 groups, seems to be slowly quenched without an intermediate starburst
 phase, resulting in the observed reduction in the mean EW(H$_\alpha$)
 values seen in the vicinity of groups in Fig.~\ref{mean_ha}. This
 observation could be attributed to the relatively inefficient
 ram-pressure stripping in galaxy groups \citep{tanaka}, transforming
 infalling galaxies slowly over several Gyr via a process involving
 progressive starvation \citep{larson}.

Another notable fact in Fig.~\ref{mean_ha} is the orientation of the
stream of blue star-forming galaxies southward of Coma, almost
orthogonal to the direction of elongation of the galaxy density
(Figs.~\ref{sf_frac} \& \ref{mean_ha}). This may be significant given
the observations presented in the previous section \citep[also
see][]{burns94}. Consistent with the results presented in
literature, we find that the instantaneous SFR of a galaxy
depends upon the stellar mass of the galaxy, as well as on the
local galaxy density, and on whether  the galaxy
is in a group or cluster.
The cosmic web does play a crucial role in defining the
evolutionary path of the galaxy, as is seen by different rate of
quenching of dwarf galaxies in groups and clusters
(Fig.~\ref{mean_ha}). The different fractions of dwarf k+A galaxies
in the three major components of this supercluster further strengthens
this argument (Table~\ref{tbl:ka-nos}). It is indeed remarkable that
the Coma and Abell~1367 clusters, that are well known to have very
different galaxy properties,  influence the star formation 
histories of the infalling  galaxies very
differently. While the fraction of star-forming and k+A dwarf galaxies
in Abell 1367 vary by a factor of 10 ($\sim$\,52\% \& 4.8\%
respectively; Table~\ref{tbl:ka-nos}), in the Coma cluster neither
dominate (12.5\% \& 11.4\% respectively; Table~\ref{tbl:ka-nos}).

 \section{Conclusions}
 \label{conclusions}

 This work is a step ahead in understanding the star formation
 properties of galaxies in one of the richest nearby superclusters. We
 analyse the spectroscopic and photometric data obtained by the SDSS
 and archival 24\m\ data obtained by the MIPS instrument on board {\em
   Spitzer}. Our major results are:

 \begin{itemize}	
 \item The fraction of (optical) AGN drops significantly
   ($f_{AGN}\!<\!0.25$) in the dense cluster environment. But the
   relation between AGN activity and environment is unclear in the
   intermediate density environments of galaxy groups.

 \item Star formation in massive galaxies ($z\!<\!14.5$) seems to
   be low everywhere in the supercluster region studied here, almost
   independent of the local environment. In sharp contrast, the
   dwarf galaxies ($z\!>\!15$) can be seen to be rapidly forming stars
   everywhere, except in the dense cluster and group environments.

 \item The passive, AGN host and star-forming galaxies as classified
   from their optical spectra, occupy different regions on the
   $(24\!-\!z)\!-\!(g\!-\!r)$ colour-colour diagram.

 \item The fraction of star-forming galaxies in Coma is determined to
   within $\pm 10$\%, irrespectively of the definition of star-forming
   in terms of optical $(g\!-\!r)$ colour, optical-mid-IR $(24\!-\!z)$
   colour or EW(H$_\alpha$).  Many of the blue galaxies in Coma are
   found to be post-starburst galaxies, whose blue colours are due to
   a recent burst of star formation which has now terminated, as
   revealed by their lack of H$_\alpha$ emission and excess H$_\delta$
   absorption.  However, in Abell~1367, the $f_{SF}$ obtained using
   the 3 different indicators show different trends. While the
   fraction of blue galaxies increases outward from the centre, the
   $f_{SF}$ obtained by employing the ($24\!-\!z$) near/mid IR colour
   decreases away from the centre of Abell 1367.

 \item Most of the (optically) blue 24\m\ galaxies detected in Coma are on their
   first infall towards the cluster. The current episode of star
   formation in such galaxies is possibly a result of a rapidly changing local
   environment.

 \item 11.4\% of all dwarf ($z\!>\!15$) galaxies within
   $5\!\times\!4.2$~$h^{-1}_{70}$~Mpc$^2$ of the centre of Coma, and
   4.8\% within the same area around Abell 1367 have post-starburst
   (k+A) type spectra. In the surrounding supercluster region this
   fraction drops to 2.1\% only, suggesting that the mechanism(s)
   responsible for quenching star formation in dwarfs depends upon the
   cluster's potential. The starburst, rapid quenching and subsequent
   k+A phase requires the dense ICM and high infall velocities attainable in
   rich clusters, as opposed to galaxy groups where star formation in
   infalling dwarf galaxies appears to be quenched gradually. The k+A
   galaxies preferentially avoid the dense centre of the cluster.

 \item The spatial distribution of the k+A galaxies suggests a
   correlation between the substructure of the Coma cluster (revealed
   in the X-ray emission) and the mechanisms responsible for quenching
   star formation in galaxies.

 \end{itemize}

 \section{Acknowledgments}

 We thank Dr Alexis Finoguenov and Dr Ulrich Briel for providing us the
 XMM-Newton EPIC/PN 0.5--2~keV mosaic image of the Coma cluster used
 in Figs.~\ref{coma-gals} and \ref{coma-mips}.  SM is supported by
 grants from ORSAS, UK, and the University of Birmingham. CPH
 acknowledges financial support from STFC.
 We are very grateful to the anonymous referee for constructive comments
 that were very helpful in improving this paper. 
 
 This research has made use of the SAO/NASA Astrophysics Data System,
 and the NASA/IPAC Extragalactic Database (NED).  
 Funding for the SDSS and SDSS-II has been provided by the Alfred
 P. Sloan Foundation, the Participating Institutions, the National Science
 Foundation, the U.S. Department of Energy, the National Aeronautics and Space
 Administration, the Japanese Monbukagakusho, the Max Planck Society, and the Higher
 Education Funding Council for England. A list of participating institutions
 can be obtained from the SDSS Web Site http://www.sdss.org/.

%%%%%%%%%%%%%% Bibiliography %%%%%%%%%%%%%%%%%%%%%%%%%%%%%%%
 
 \label{lastpage}

 \end{document}